\documentclass[a4paper,11pt]{article}
\usepackage{jheppub}
\usepackage{amsthm}
\usepackage{amssymb}
\usepackage{amsthm}
\usepackage{amstext}
\usepackage{amsmath}
\usepackage{breqn}
\usepackage{epstopdf}
\setkeys{breqn}{breakdepth={0}}
\usepackage{amsfonts}
\usepackage{bbm}
\usepackage[german,english]{babel}

\usepackage{verbatim}

\usepackage{caption,subcaption}
\usepackage{graphicx} 
\usepackage{epstopdf}
\usepackage{float}

\usepackage{paralist}
\usepackage{algorithm}
\usepackage{algorithmic}

\usepackage{tikz}
\usetikzlibrary{positioning}
\usetikzlibrary{arrows.meta}

\usepackage{listings}
\usepackage{xcolor}
\usepackage{todonotes}
\definecolor{codegreen}{rgb}{0,0.6,0}
\definecolor{codegray}{rgb}{0.5,0.5,0.5}
\definecolor{codepurple}{rgb}{0.58,0,0.82}
\definecolor{backcolour}{rgb}{0.93,0.93,0.93}

\newcommand{\dif}{\mathrm{d}}

\theoremstyle{plain}

\theoremstyle{definition}

\allowdisplaybreaks[4]

\title{A nice two-loop next-to-next-to-MHV amplitude in ${\cal N}=4$ super-Yang-Mills}

\author[a,b,c]{Song He}\author[a]{Zhenjie Li}\author[d]{Chi Zhang}

\affiliation[a]{CAS Key Laboratory of Theoretical Physics, Institute of Theoretical Physics, Chinese Academy of Sciences, Beijing 100190, China}
\affiliation[b]{School of Fundamental Physics and Mathematical Sciences, Hangzhou Institute for Advanced Study \&  ICTP-AP, UCAS, Hangzhou 310024, China}
\affiliation[c]{Peng Huanwu Center for Fundamental Theory, Hefei, Anhui 230026, China}
\affiliation[d]{%
Niels Bohr International Academy, Niels Bohr Institute, Copenhagen University, Blegdamsvej 17, 2100 Copenhagen \O{}, Denmark}

\abstract{We study a scalar component of the 8-point next-to-next-to-maximally-helicity-violating (N${}^2$MHV) amplitude at two-loop level in ${\cal N}=4$ super-Yang-Mills theory; it has a leading singularity proportional to the inverse of the four-mass-box square root and receives contributions from only two types of non-trivial integrals with one-loop infrared (IR) divergences. We compute such two-loop 8-point integrals by taking (double-)collinear limits of certain finite, dual-conformal-invariant integrals, and they nicely give the IR-safe ratio function after subtracting divergences. As the first genuine two-loop N${}^2$MHV amplitude computed explicitly, we find remarkable structures in its symbol and alphabet: similar to the next-to-MHV (NMHV) case, there are still $9$ algebraic letters associated with the square root, and the latter also becomes a letter for the first time; unlike the NMHV case, such algebraic letters appear at either one or all of the second, third and last entry, and the part with three odd letters is particularly simple.}

\emailAdd{songhe@mail.itp.ac.cn, lizhenjie@itp.ac.cn, chi.zhang@nbi.ku.dk}

\begin{document}

\maketitle
\section{Introduction}

Scattering amplitudes are central objects in fundamental physics: they are crucial for connecting theory to experiments in particle accelerators such as Large Hadron Collider, and they play a central role in discovering new structures of Quantum Field Theory (QFT). In particular, tremendous progress has been made for planar ${\cal N} = 4$ supersymmetric Yang-Mills theory (SYM); not only have hidden mathematical structures for all-loop integrands been unraveled~\cite{ArkaniHamed:2010kv, Arkani-Hamed:2012zlh, Arkani-Hamed:2013jha}, but amplitudes have also been determined to impressively high loops, for $n = 6, 7$~\cite{Dixon:2011pw, Dixon:2014xca, Dixon:2014iba, Drummond:2014ffa, Dixon:2015iva, Caron-Huot:2016owq,Dixon:2016nkn, Drummond:2018caf, Caron-Huot:2019vjl, Caron-Huot:2019bsq, Dixon:2020cnr} and beyond~\cite{Anastasiou:2009kna, CaronHuot:2011ky, Zhang:2019vnm, He:2020vob}. 
A remarkable duality between Maximally Helicity Violating (MHV) scattering amplitudes and null polygonal Wilson loops (WL) of the theory was discovered at both strong~\cite{Alday:2007hr, Alday:2007he, Alday:2009yn} and weak coupling~\cite{Brandhuber:2007yx,Drummond:2007aua,Drummond:2007cf,Drummond:2007bm,Drummond:2008aq, Bern:2008ap}; later it was established that super-amplitudes (after stripping off MHV tree prefactor) are dual to supersymmetric WL~\cite{CaronHuot:2010ek, Mason:2010yk}, and quite a lot of what we have learned about amplitudes are from this duality picture. For example, one can compute amplitudes at any value of the coupling around collinear limits~\cite{Basso:2013vsa} based on integrability~\cite{Beisert:2010jr} and operator product expansions (OPE) of WL~\cite{Alday:2010ku}. From such a dual WL picture, one can derive the powerful $\bar{Q}$ anomaly equation~\cite{CaronHuot:2011kk}, which has been the driving force for computing multi-loop amplitudes, including two-loop MHV to all multiplicities~\cite{Caron-Huot:2011zgw}, $n\leq 9$ NMHV~\cite{CaronHuot:2011kk, Zhang:2019vnm,He:2020vob} and even three-loop MHV for $n\leq 8$~\cite{Li:2021bwg}. 

As shown in~\cite{CaronHuot:2011kk}, given lower-loop amplitudes with higher $n,k$, the ${\bar Q}$ anomaly equation can be used to {\it uniquely} determine MHV and NMHV amplitudes, up to ``constants" which are in the kernel of $\bar{Q}$ operator: any function that satisfies $\bar{Q} F_{n,k}=0$ must be a constant for $k=0$ or constant times R-invariants for $k=1$~\cite{CaronHuot:2011kk}. However, this is no longer true for $k\geq 2$: there exist transcendental functions in the kernel of ${\bar Q}$ for N${}^2$MHV amplitudes and beyond. In principle such amplitudes can be determined by exploiting anomaly equations for level-one generators such as $Q^{(1)}$, which can be obtained from spacetime parity-conjugation of the $\bar{Q}$ equation, but it is a formidable task already for the simplest case with $n=8, k=2$. It remains an important open question to determine, or constrain as much as possible, multi-loop amplitudes with $k\geq 2$ using ${\bar Q}$ and $Q^{(1)}$ anomaly equations; in particular, what do these equations say about amplitudes which evaluate to functions beyond multiple polylogarithms (MPL) such as $n=10, k=3$ case~\cite{Caron-Huot:2012awx,Bourjaily:2017bsb,Kristensson:2021ani}?

In this paper, we will take the more traditional approach, namely directly computing Feynman loop integrals contributing to (certain components of) amplitudes with $k\geq 2$, which has proved very successful for {\it e.g.} two-loop MHV and NMHV amplitudes. The main motivation is to obtain more data about scattering amplitudes and Feynman integrals in ${\cal N}=4$ SYM, which in turn allows us to discover more mathematical structures about their symbology and physics related to the ${\bar Q}$ anomaly equation {\it etc.}. Our main interests are two-loop N${}^2$MHV amplitudes, where the integrands in four-dimensional space are known from generalized unitarity~\cite{Bourjaily:2015jna}; at least for $n=8$, all such integrals are expected to evaluate to MPLs, which can in principle be done by using direct integration method~\cite{CaronHuot:2011kk} if infrared divergences are suitably regulated. We will not compute the entire two-loop 8-point N${}^2$MHV super-amplitude through this method, mainly for two reasons: (1) there are too many integrals with rather complicated kinematic dependence with as many as $9$ dual-conformal-invariant cross-ratios; additional technical difficulties are caused by regulating divergent integrals; (2) we expect that most components of N${}^2$MHV amplitudes are in a sense similar to those of NMHV/MHV amplitudes, since they receive contributions from same types of integrals and there are certain similarities regarding their symbology; thus we will focus on what we consider as ``genuinely new" components of N${}^2$MHV amplitudes, which exhibit new structures invisible at NMHV/MHV level. 

Therefore, we will focus on a class of particularly nice components of N${}^2$MHV amplitudes,
which vanish at tree level and remain finite at one-loop level. The (algebraic, Grassmann-valued) coefficients of these integrals, which are leading singularities or Yangian invariants, are well understood: for $k=2$ only one special class of them are non-rational functions with square roots, which first appear in one-loop amplitudes as coefficients of four-mass box integrals.  Certain scalar components isolate such a non-rational Yangian invariant~\cite{ArkaniHamed:2009dn,Arkani-Hamed:2012zlh}, thus {\it e.g.} at one loop it is simply given by a finite four-mass box integral which contains the same square root. For $n=8$, there are exactly two such scalar components related to each other by a cyclic rotation, and {\it e.g.} one of them, 
\[
\mathcal{A}(\varphi_{12},\varphi_{12},\varphi_{13},\varphi_{13},\varphi_{34},\varphi_{34},\varphi_{24},\varphi_{24})\:,
\]
which depends on the square root $\Delta_{1,3,5,7}$ is given by the four-mass box integral with dual points $x_1, x_3, x_5, x_7$, which will be denoted as $\Delta$ when there is no ambiguity. At two-loop level, as we will see shortly, such a component receives contributions from three integrals (up to cyclic rotations), a penta-box and a double-box, both of which has only one-loop divergence, and the well-known finite box ladder integral~\cite{Usyukina:1993ch}; the finite {\it ratio function} can be obtained by subtracting one-loop MHV amplitudes multiplied by the one-loop component (which is given by the four-mass box function). We will use the dual conformal invariant (DCI) regularization~\cite{Bourjaily:2013mma} since these integrands are defined strictly in four-dimensional space: it turns out the $n=8$ penta-box and double-box can be obtained from double-collinear limits of $n=10$ finite DCI integrals, and the DCI regularization allows us to extract these divergent integrals from the latter: by sending a dual point to infinity for the $n=10$ penta-box integral, it is nicely reduced to some two-loop master integrals with four masses which have been recently computed very recently using differential equations~\cite{He:2022ctv}; the $n=10$ double-box evaluates to elliptic multiple polylogarithms \cite{Kristensson:2021ani}, but we will see that it suffices to take a double-collinear limit for its one-dimensional integral representation, which gives the $n=8$ double-box as MPLs. After subtracting the DCI-regulated one-loop 8-point MHV amplitude multiplied by the four-mass box function, we confirm the cancellation of divergences and end up with a finite ratio function. 

Having computed these cutting-edge examples for the two-loop DCI-regulated integrals contributing to this N${}^2$MHV component, we will then study the structures of the resulting symbol and the alphabet, especially in comparison with earlier results of $n=8$ NMHV and MHV amplitudes~\cite{Caron-Huot:2011zgw,Zhang:2019vnm,Li:2021bwg}. We find that the alphabet of this component (and its cyclic rotation) stays very similar to the NMHV one (MHV case is absent of any square root), which consists of $180$ rational letters and $9+9$ algebraic letters (associated with square root $\Delta$ and its cyclic rotation)~\cite{Zhang:2019vnm} (the appearance of these algebraic letters is studied from many different directions~\cite{He:2020uhb,Mago:2020kmp,Arkani-Hamed:2019rds,Drummond:2019cxm,Drummond:2019qjk,Drummond:2020kqg,Henke:2019hve,Henke:2021ity,He:2021eec,He:2021esx,He:2021non,Yang:2022gko}); in particular there are still $9$ independent algebraic letters which are {\it odd} under the flip $\Delta \to -\Delta$ (all rational letters are even), and there is exactly one new letter for this N${}^2$MHV component, \emph{namely $\Delta$ itself}. Furthermore, recall that NMHV amplitudes or other components of N${}^2$MHV ones must be even under the flip, thus at two loops they contain none or two odd (algebraic) letters, which can only be in the second and third entry~\cite{Zhang:2019vnm,He:2020vob} (some individual Feynman integrals also have this property~\cite{He:2020uxy,He:2020lcu}); in particular this means that the last entry only consists of even (rational) letters, as predicted by $\bar{Q}$ equation. On the contrary, this special component of N${}^2$MHV must be {\it odd}, and at two loops it contains either one or three odd letters, in the second, third and last entries. Among other things, we find remarkably simple results for the part containing three odd letters.

\section{A nice component of 8-point N${}^2$MHV: two-loop ratio function}
\label{sec:integrand}
A nice representation for two-loop integrands of  $n$-point, N${}^k$MHV super-amplitude in ${\cal N}=4$ SYM has been obtained using {\it generalized unitarity}~\cite{Bourjaily:2015jna}. Schematically, the amplitudes in planar ${\cal N}=4$ SYM take the form
\begin{equation} \label{generalized_unitarity_amp}
    \mathcal{A}_{n,k} = \sum \text{Leading Singularity}  \times \text{Scalar Integral} 
\end{equation}
where leading singularities are Yangian invariants: they are completely fixed by super conformal symmetries and dual super conformal symmetries and hence independent of loop level~\cite{Arkani-Hamed:2012zlh}.

For our purpose, it is convenient to work with the \emph{dual} superspace coordinates $(x,\theta)$,
\begin{equation}
    x_{i+1}^{\alpha\dot{\alpha}}-x_{i}^{\alpha\dot{\alpha}}=p_{i}^{\mu}\sigma_{\mu}^{\alpha\dot{\alpha}}=\lambda_{i}^{\alpha}\tilde{\lambda}_{i}^{\dot{\alpha}}\:, \qquad \theta_{i+1}^{\alpha A}-\theta_{i}^{\alpha A}=\lambda_{i}^{\alpha}\eta_{i}^{A},
\end{equation}
as well as the following shorthand notation for the Lorentz invariants,
\[
x_{i,j}^{2}=(x_{i}-x_{j})^{2}=(p_{i}+p_{i+1}+\cdots +p_{j-1})^{2}.
\]
To linearly realize the dual superconformal symmetries and have nice expressions for Yangian invariants, it is convenient to introduce (super) momentum twistor variables~\cite{Hodges:2009hk},
\begin{equation}
    \mathcal{Z}_{i}=(Z_{i}^{a}\vert\chi_{i}^{A}):=(\lambda_{i}^{\alpha},x_{i}^{\alpha\dot{\alpha}}\lambda_{i\alpha}\vert \theta_{i}^{\alpha A}\lambda_{i \alpha})\:.
\end{equation}
In terms of super momentum twistors, we further define the basic $\mathrm{SL}(4)$-invariant $\langle ijkl\rangle:=\epsilon_{abcd}Z_{i}^{a}Z_{j}^{b}Z_{k}^{c}Z_{l}^{d}$ (or the Pl\"{u}cker coordinates of $\mathrm{Gr}(4,n)$), and the basic $R$ invariant~\cite{Drummond:2008vq,Mason:2009qx}
\begin{equation}\label{Rinv}
    [i\,j\,k\,l\,m]:=\frac{\delta^{0\vert 4}(\chi_{i}^{A}\langle jklm\rangle+\text{cyclic})}{\langle ijkl\rangle\langle jklm\rangle
    \langle klmi\rangle\langle lmij\rangle\langle mijk\rangle} \:.
\end{equation}

The genuine N$^{2}$MHV amplitudes in planar $\mathcal{N}=4$ SYM first appear starting from $n=8$, which is sufficient for our purpose. As we have mentioned, we are mostly interested in (two cyclically related) component amplitudes, which are proportional to Yangian invariants with square roots; such a scalar component reads:
\begin{align}
\mathcal{A}(\varphi_{12},\varphi_{12},\varphi_{13},\varphi_{13},\varphi_{34},\varphi_{34},\varphi_{24},\varphi_{24})&=\int \dif\chi_{1}^{1}\dif\chi_{2}^{1}\dif\chi_{3}^{2}\dif\chi_{4}^{2}\chi_{5}^{3}\dif\chi_{6}^{3}\dif\chi_{7}^{4}\dif\chi_{8}^{4}\: {\cal A}_{8,2} \nonumber \\
&=\sum_{L=0}^{\infty} (g^2)^L \mathbf{A}_{8,2}^{(L)}\,,
\end{align}
where $\mathbf{A}_{8,2}^{(L)}$ denotes the $L$-loop contribution to this component amplitude. 
At tree level, this component amplitude vanishes, $\mathbf{A}_{8,2}^{(0)}=0$, the one-loop contribution $\mathbf{A}_{8,2}^{(1)}$ thus is finite. Moreover, $\mathbf{A}_{8,2}^{(1)}$ receives contribution only from one term in eq.\eqref{generalized_unitarity_amp},
\begin{equation}
  \sum_{\pm} \frac{(1-u-v\pm\Delta)}{2\Delta}[\alpha_{\pm},4,5,6,7][\gamma_{\pm},8,1,2,3] \int \frac{-\Delta \,x_{1,5}^{2} x_{3,7}^{2}\:\dif^{4} x_{0}}{x_{1,0}^{2}x_{3,0}^{2}x_{5,0}^{2}x_{7,0}^{2}}
\end{equation}
where $\alpha$ and $\delta$ are two solutions of the Schubert problem
\begin{equation}
    \alpha = (a{-}1 \, a) \cap (d\,d{-}1\,\gamma) \:, \qquad 
    \gamma = (c{-}1\,c) \cap (b\,b{-}1,\alpha) \:,
\end{equation}
and 
\begin{equation}
  u = \frac{x_{1,3}^{2}x_{5,7}^{2}}{x_{1,5}^{2}x_{3,7}^{2}}\:, \quad 
  v = \frac{x_{1,7}^{2}x_{3,5}^{2}}{x_{1,5}^{2}x_{3,7}^{2}}\:, \quad  \Delta= \sqrt{(1-u-v)^{2}-4u v} \:.
\end{equation}
By integrating Grassmannian variables, we have the scalar component $\mathbf{A}_{8,2}^{(1)}$, 
\begin{align}
    \mathbf{A}_{8,2}^{(1)}&=I^{(1)}_{\rm bl}:=\int \frac{\dif^{4} x_{0}}{x_{01}^{2}x_{03}^{2}x_{05}^{2}x_{07}^{2}} \nonumber
    \\
    &=\frac{2}{x_{1,5}^2x_{3,7}^2\Delta}\biggl(\text{Li}_2(z)-\text{Li}_2(\bar z)+\frac{1}{2} \log (z \bar z) \log \left(\frac{1-z}{1-\bar z}\right)\biggr), \label{boxInt}
\end{align}
where the box integral has been denoted as the one-loop instance of well-known box ladders~\cite{Usyukina:1993ch},  $I^{(1)}_{\rm bl}$, and $z, \bar{z}, 1-z, 1-\bar{z}$, which contain the square root $\Delta$, are defined by
\[
z \bar z=u,\quad (1-z)(1-\bar z)=v.
\]

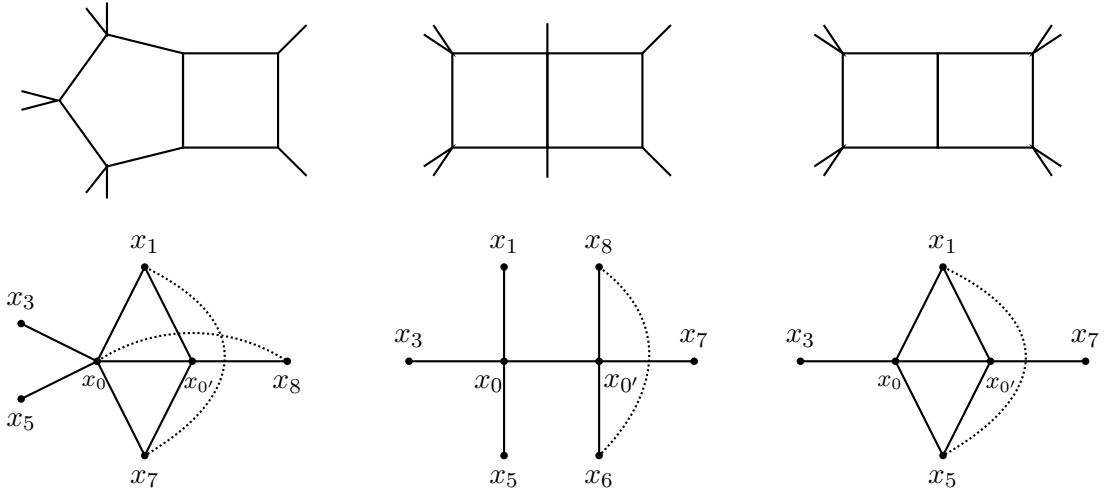
\begin{figure}
\hspace{0.42cm}
\begin{tikzpicture}[baseline={([yshift=-.5ex]current bounding box.center)},scale=0.25]
    \draw[black,thick] (-9,6)--(-5,5) -- (-5,0) --(-9,-1)--(-11.5,2.5)--cycle;
   \draw[black,thick] (-10.1,7.4)--(-9, 6)--(-9.0, 7.7);
    \draw[black,thick] (-10.1,-2.4)--(-9, -1)--(-9.0,-2.7);
    \draw[black,thick] (-5,0)--(-5,5)--(0,5)--(0,0)--cycle;
    \draw[black,thick] (1.5,6.5)--(0,5);
    \draw[black,thick] (1.5,-1.5)--(0,0);
    \draw[black,thick] (-13.5,3)--(-11.6,2.5)--(-13.5,2);
\end{tikzpicture} \hspace{1.26cm}
\begin{tikzpicture}[baseline={([yshift=-.5ex,xshift=-1cm]current bounding box.center)},scale=0.25]
    \draw[black,thick] (0,5)--(-5,5)--(-5,0)--(0,0)--cycle;
    \draw[black,thick] (-5,5) -- (-5,6.6);
    \draw[black,thick] (-5,0) -- (-5,-1.6);
    \draw[black,thick] (-11.5,6)--(-10,5)--(-11,6.5);
    \draw[black,thick] (-11,-1.5)--(-10,0)--(-11.5,-1);
    \draw[black,thick] (-5,0)--(-5,5)--(-10,5)--(-10,0)--cycle;
    \draw[black,thick] (0,5)--(1.5,6.5);
    \draw[black,thick] (1.5,-1.5)--(0,0);
\end{tikzpicture} \hspace{1.60cm}
\begin{tikzpicture}[baseline={([yshift=-.5ex]current bounding box.center)},scale=0.25]
    \draw[black,thick] (0,5)--(-5,5)--(-5,0)--(0,0)--cycle;
    \draw[black,thick] (-11.5,6)--(-10,5)--(-11,6.5);
    \draw[black,thick] (-11,-1.5)--(-10,0)--(-11.5,-1);
    \draw[black,thick] (-5,0)--(-5,5)--(-10,5)--(-10,0)--cycle;
    \draw[black,thick] (1.5, 6)--(0,5)--(1,6.5);
    \draw[black,thick] (1,-1.5)--(0,0)--(1.5,-1);
\end{tikzpicture} 
\begin{center}
    \begin{tikzpicture}[baseline={([yshift=-.5ex]current bounding box.center)},scale=0.5]
        \draw[thick] (-0.5,-2.5) -- (1.5,-3.5);
        \draw[thick] (-0.5,-4.5) -- (1.5,-3.5);
        \draw[thick] (2.75,-1) -- (1.5,-3.5);
        \draw[thick] (2.75,-6) -- (1.5,-3.5);
        \draw[thick] (4,-3.5) -- (1.5,-3.5);
        \draw[thick] (4,-3.5) -- (6.5,-3.5);
        \draw[thick] (4,-3.5) -- (2.75,-6);
        \draw[thick] (4,-3.5) -- (2.75,-1);
        \draw[thick,densely dotted] (1.5,-3.5) .. controls (3,-2.5) and (5,-2.5) .. (6.5,-3.5);
        \draw[thick,densely dotted] (2.75,-1) .. controls (5.8,-2.5) and (5.3,-4.5) .. (2.75,-6);
        \node[fill=black,circle,inner sep=1pt,label=below:$x_{5}$] at (-0.5,-4.5) {};
        \node[fill=black,circle,inner sep=1pt,label=below:$x_{8}$] at (6.5,-3.5) {};
        \node[fill=black,circle,inner sep=1pt,label={[xshift=0.1cm, yshift=-0.6cm]:{\footnotesize $ x_{0'}$}}] at (4,-3.5) {};
        \node[fill=black,circle,inner sep=1pt,label=below:$x_{7}$] at (2.75,-6) {};
        \node[fill=black,circle,inner sep=1pt,label=below:{\footnotesize $ x_{0}\:$}] at (1.5,-3.5) {};
        \node[fill=black,circle,inner sep=1pt,label=above:$x_{1}$] at (2.75,-1) {};
        \node[fill=black,circle,inner sep=1pt,label=above:$x_{3}$] at (-0.5,-2.5) {};
    \end{tikzpicture} \hspace{0.7cm}
    \begin{tikzpicture}[baseline={([yshift=-.5ex]current bounding box.center)},scale=0.5]
    \draw[thick] (-1,-3.5) -- (1.5,-3.5);
    \draw[thick] (1.5,-1) -- (1.5,-3.5);
    \draw[thick] (1.5,-6) -- (1.5,-3.5);
    \draw[thick] (4,-3.5) -- (1.5,-3.5);
    \draw[thick] (4,-3.5) -- (6.5,-3.5);
    \draw[thick] (4,-3.5) -- (4,-6);
    \draw[thick] (4,-3.5) -- (4,-1);
    \draw[thick,densely dotted] (4,-1) .. controls (6,-2.5) and (5.5,-4.5) .. (4,-6);
    \node[fill=black,circle,inner sep=1pt, label=above:$x_{3}$] at (-1,-3.5) {};
    \node[fill=black,circle,inner sep=1pt, label=above:$x_{7}$] at (6.5,-3.5) {};
    \node[fill=black,circle,inner sep=1pt,label=above:$x_{8}$] at (4,-1) {};
    \node[fill=black,circle,inner sep=1pt,label={[xshift=0.3cm, yshift=-0.6cm]:{$ x_{0'}$}}] at (4,-3.5) {};
    \node[fill=black,circle,inner sep=1pt,label=below:$x_{6}$] at (4,-6) {};
    \node[fill=black,circle,inner sep=1pt,label=below:$x_{5}$] at (1.5,-6) {};
    \node[fill=black,circle,inner sep=1pt,label={[xshift=-0.2cm, yshift=-0.6cm]:{$ x_{0}$}}] at (1.5,-3.5) {};
    \node[fill=black,circle,inner sep=1pt,label=above:$x_{1}$] at (1.5,-1) {};
\end{tikzpicture} \hspace{0.5cm}
     \begin{tikzpicture}[baseline={([yshift=-.5ex]current bounding box.center)},scale=0.5]
        \draw[thick] (-1,-3.5) -- (1.5,-3.5);
        \draw[thick] (2.75,-1) -- (1.5,-3.5);
        \draw[thick] (2.75,-6) -- (1.5,-3.5);
        \draw[thick] (4,-3.5) -- (1.5,-3.5);
        \draw[thick] (4,-3.5) -- (6.5,-3.5);
        \draw[thick] (4,-3.5) -- (2.75,-6);
        \draw[thick] (4,-3.5) -- (2.75,-1);
        \draw[thick,densely dotted] (2.75,-1) .. controls (5.9,-2.5) and (5.4,-4.5) .. (2.75,-6);
        \node[fill=black,circle,inner sep=1pt,label=above:$x_{3}$] at (-1,-3.5) {};
        \node[fill=black,circle,inner sep=1pt,label=above:$x_{7}$] at (6.5,-3.5) {};
        \node[fill=black,circle,inner sep=1pt,label={[xshift=0.15cm, yshift=-0.6cm]:{\footnotesize $ x_{0'}$}}] at (4,-3.5) {};
        \node[fill=black,circle,inner sep=1pt,label=below:$x_{5}$] at (2.75,-6) {};
        \node[fill=black,circle,inner sep=1pt,label=below:{\footnotesize $ x_{0}\:\:$}] at (1.5,-3.5) {};
        \node[fill=black,circle,inner sep=1pt,label=above:$x_{1}$] at (2.75,-1) {};
    \end{tikzpicture}
\end{center}
\caption{Three Feynman diagrams as well as their dual diagrams that contribute to the two-loop N${}^{2}$MHV component amplitude $\mathbf{A}^{(2)}_{8,2}$: the penta-box diagram, the double-box diagram, and the box-ladder diagram. In the dual diagrams, the line connecting to two dual points, $x_{a}$ and $x_{b}$, represents a factor $x_{a,b}^{2}$ in the denominator, while the dotted line represents a factor in the numerator.}
\end{figure}

At two-loop level, by repeating the procedure as in the one-loop case, one can find only three types of Feynman integrals that contribute to this component amplitude: up to cyclic rotations, we have a penta-box integral with $5$ dual points, $I_{\rm pb}(x_1, x_3, x_5, x_7,x_{8})$, a double-box integral with $6$ dual points, $I_{\rm db}(x_1, x_3, x_5,x_{6}, x_7,x_{8})$, and the double box integral with $4$ dual points $I_{\rm bl}^{(2)} (x_1, x_3, x_5, x_7)$ ($L=2$ instance of the box ladder $I_{\rm bl}^{(L)}$)~\cite{Usyukina:1993ch}:
\begin{equation}
\mathbf{A}_{8,2}^{(2)}=\left(2I_{\rm pb}(x_1, x_3, x_5, x_7, x_8) + 2I_{\rm db}(x_1, x_3, x_5, x_6, x_7, x_8)+I_{\rm bl}^{(2)} (x_1, x_3, x_5, x_7)\right) + 3~{\rm cyclic}
\end{equation}
where these three types of Feynman integrals read
\begin{align}
I_{\rm pb}(x_1, x_3, x_5, x_7,x_{8})&=\int \frac{  x_{8,0}^{2} x_{1,7}^{2} \: \dif^{4} x_{0} \dif^{4} x_{0'}}{ x_{1,0}^{2} x_{3,0}^{2} x_{5,0}^{2} x_{7,0}^{2} x_{0,0'}^{2} x_{7,0'}^{2} x_{8,0'}^{2} x_{1,0'}^{2}} \:, \label{pbintegrand} \\
I_{\rm db}(x_1, x_3, x_5,x_{6}, x_7,x_{8})&=\int \frac{ x_{6,8}^{2}\: \dif^{4} x_{0} \dif^{4}x_{0'} }{x_{1,0}^{2}x_{3,0}^{2}x_{5,0}^{2}x_{0,0'}^{2}x_{6,0'}^{2} x_{7,0'}^{2}x_{8,0'}^{2}} \:, \label{dbintegrand} \\
I_{\rm bl}^{(2)}(x_1, x_3, x_5, x_7)&=\int \frac{x_{1,5}^{2}\: \dif^{4} x_{0} \dif^{4} x_{0'}}{x_{1,0}^{2}x_{3,0}^{2}x_{5,0}^{2}x_{0,0'}^{2}x_{5,0'}^{2}x_{7,0'}^{2}x_{1,0'}^{2}} \:, \label{blintegrand}
\end{align}
the factor `2' in front of $I_{\rm pb}$ and $I_{\rm db}$ is due to their symmetry under the exchange $x_{0}\leftrightarrow x_{0'}$, `$+3$ cyclic' means taking $x_i\to x_{i{+}2}$ for $i=1,\cdots, 8$ which gives a cyclic orbit of length $4$. As remarked in the introduction and shown in the one-loop case eq.\eqref{boxInt}, this component amplitude $\mathbf{A}_{8,2}^{(2)}$ as well as its three ingredient integrals are pure polylogarithms divided by the square root 
\[
x_{1,5}^{2}x_{3,7}^{2}\Delta= \sqrt{(x_{1,5}^{2}x_{3,7}^{2}-x_{1,3}^{2}x_{5,7}^{2}-x_{3,5}^{2}x_{1,7}^{2})^{2}-4x_{1,3}^{2}x_{3,5}^{2}x_{5,7}^{2}x_{1,7}^{2}} \:,
\]
we thus introduce the normalized amplitudes and Feynman integrals indicated by a hat:
\begin{equation} \label{pure_integrals_amplitudes}
    \widehat{\mathbf{A}}_{8,2} = x_{1,5}^{2}x_{3,7}^{2}\Delta  \times \mathbf{A}_{8,2} \:, \quad 
    \widehat{I}_{\bullet}=x_{1,5}^{2}x_{3,7}^{2}\Delta \times I_{\bullet} \:. 
\end{equation}

This component suffers from infrared divergence: both the penta-box integral $I_{\rm pb}$ and double-box integral $I_{\rm db}$ has one-loop divergence from the box with two massless legs ($x_{0'}$ in the above equations). However, it is well known that the infrared divergence is captured by the MHV amplitudes (also known as BDS ansatz) \cite{Bern:2005iz},  then it is clear that the so-called ratio functions $\mathcal{R}_{n,k}=\mathcal{A}_{n,k}/\mathcal{A}_{n,0}$ are finite and independent of the regularization scheme. For our case, the corresponding component of $\mathcal{R}_{8,2}$ is slightly simpler and reads
\begin{equation} \label{ratio_function_component}
    \mathbf{R}_{8,2}^{(2)} =\mathbf{A}^{(2)}_{8,2}-\mathbf{A}^{(1)}_{8,2} \times \mathcal{A}_{8,0}^{(1)}
\end{equation}
due to the vanishing $\mathbf{A}_{8,2}^{(0)}$. As above, we introduce the normalized (pure) ratio function $\widehat{\mathbf{R}}_{8,2}$. The infrared structure of $\mathbf{A}_{8,2}^{(2)}$ now is manifest: it is captured by the one-loop 8-point MHV amplitude $\mathcal{A}_{8,0}^{(1)}$. Note that this component ratio function is the same as the BDS-normalized amplitude $(\mathcal{A}_{n,k}/\mathcal{A}_{\text{BDS}})^{(2)}$ at two-loop order. 

The general regularization for the Feynman integrals in Gauge theories is dimensional regularization which will spoil the dual conformal symmetries of $\mathcal{N}=4$ SYM, we instead use the so-called \emph{dual-conformal regularization}~\cite{Bourjaily:2013mma}, which will be elaborated in the following section.




\section{The computation: DCI-regulated penta-box and double-box integrals and the subtraction}

The infrared divergences of the penta-box integral and the double-box integral arise from massless legs of the box sub-integral (on the right): for the penta-box, $x_{7,8}^{2}$ and $x_{8,1}^{2}$ vanish, and so do $x_{6,7}^{2}$ and $x_{7,8}^{2}$ for the double-box. To regulate these divergences, the most straightforward way is to assign a small mass $\mu$ to these massless legs, then take the limit $\mu\to 0$, which is also known as the Higgs regularization \cite{Alday:2009zm,Henn:2010ir}. In the dual space, this corresponds to a shift of dual points parameterized by the dimensional scale $\mu$ which breaks the dual conformal symmetries. The dual conformal regularization~\cite{Bourjaily:2013mma,Bourjaily:2019vby} instead shifts each dual point (for $a=1, 2, \cdots, n$) by
\begin{equation}
    x_{a} \to \widehat{x}_{a} =  x_{a}+ \epsilon (x_{a+1}-x_{a}) \frac{x_{a-2,a}^{2}}{x_{a-2,a+1}^{2}}\, \label{DCI-regulator}
\end{equation}
which is parameterized by a dimensionless parameter $\epsilon$. With this regularization, the one-loop $n$-point MHV amplitudes~\cite{Bourjaily:2013mma,Bourjaily:2019vby} read 
\begin{equation}\label{An0}
    \mathcal{A}_{n,0}^{(1)}= 
    n \log^{2}(\epsilon) + \log (\epsilon) \, \Biggl(\sum_{a=1}^{n} \log \frac{x_{a,a+2}^{2}x_{a+3,a+5}^{2}}{x_{a,a+3}^{2}x_{a+2,a+5}^{2}}\Biggr)+n\zeta_{2}+ F_{n,\text{DCI}}^{(1)} + O(\epsilon)
\end{equation}
where the finite weight-$2$ functions read
\begin{equation}
F_{n,\text{DCI}}^{(1)} = \left[
\sum_{b=4}^{\lfloor n/2 \rfloor+1} \mathrm{Li}_{2}(1-u_{1,b})+ \tfrac{1}{2}\log u_{1,b}\log v_{1,b}
\right] + \underset{\footnotesize\text{(delete duplicates)}}{\text{cyclic}} \:,
\end{equation}
with dual conformal cross-ratios
\begin{equation}
 u_{1,b}= \frac{x_{3,b}^{2}x_{2,b+1}^{2}}{x_{2,b}^{2}x_{3,b+1}^{2}}\:, \qquad 
 v_{1,b}= \frac{x_{2,n}^{2}x_{1,3}^{2}x_{b-1,b+1}^{2}x_{b,b+2}^{2}}{x_{3,n}^{2}x_{1,b}^{2}
 x_{b-1,b+2}^{2}x_{2,b+1}^{2}} \:.
\end{equation}
Similarly, one can compute one-loop $n$-point N${}^k$MHV amplitudes $\mathcal{A}_{n,k}^{(1)}$ with this regularization, and obtain the finite {\it ration function} at one-loop level: $\mathcal{R}_{n,k}^{(1)}:=\mathcal{A}_{n,k}^{(1)}-\mathcal{A}_{n,k}^{(0)} \mathcal{A}_{n,0}^{(1)}$, where $\log^2 \epsilon$ and $\log \epsilon$ terms cancel nicely~\cite{Caron-Huot:2013vda}. 

Moving to our two loop computation, in eq.\eqref{ratio_function_component} we also need to regulate the penta-box integral and the double-box integral with one-loop divergences.  Their results in dual conformal regularization can be obtained as special double-collinear limits of their finite counterparts with $n=10$, which are already known from the literature~\cite{Kristensson:2021ani,He:2022ctv}. In the rest of this section, we will elaborate the procedure of taking such collinear limits.

\subsection*{The double-box integral}

The finite counterpart of the 8-point double-box integral is the 10-point elliptic double-box integral~\cite{Bourjaily:2017bsb,Kristensson:2021ani}:
\begin{align} \label{elldbIntegrand}
    I_{\text{ell-db}}=
\begin{aligned}
\begin{tikzpicture}[scale=0.85,label distance=-1mm]
		\node[label=left:$2$] (0) at (0, 15.5) {};
		\node[label=above:$10$] (1) at (1.5, 16) {};
		\node[label=above:$1$] (2) at (0.5, 16) {};
		\node[label=above:$9$] (3) at (2.5, 16) {};
		\node[label=right:$8$] (4) at (3.0, 15.5) {};
		\node[label=below:$4$] (5) at (0.5, 14) {};
		\node[label=left:$3$] (6) at (0, 14.5) {};
		\node[label=below:$5$] (7) at (1.5, 14) {};
		\node[label=right:$7$] (8) at (3, 14.5) {};
		\node[label=below:$6$] (9) at (2.5, 14) {};
        \node(11) at (5, 15) {};
		\node (12) at (6, 15) {};
		\draw[thick] (0.center) to (4.center);
		\draw[thick] (6.center) to (8.center);
		\draw[thick] (2.center) to (5.center);
		\draw[thick] (1.center) to (7.center);
		\draw[thick] (3.center) to (9.center);
        \node[label=left:\textcolor{blue!50}{$x_3$}] (10) at (0.2, 15) {};
        \node[label=above:\textcolor{blue!50}{$x_{1}$}] (13) at (1.0, 15.8) {};
        \node[label=above:\textcolor{blue!50}{$x_{10}$}] (14) at (2.0, 15.8) {};
        \node[label=right:\textcolor{blue!50}{$x_8$}] (15) at (2.8, 15) {};
        \node[label=below:\textcolor{blue!50}{$x_6$}] (16) at (2.0, 14.3) {};
        \node[label=below:\textcolor{blue!50}{$x_5$}] (17) at (1.0, 14.3) {};
		 \draw[very thick,blue!50] (10.center) to (15.center);
		 \draw[very thick,blue!50] (13.center) to (17.center);
		 \draw[very thick,blue!50] (14.center) to (16.center);
\end{tikzpicture} 
\end{aligned}
\hspace{-0.14\textwidth}
=\int \frac{x_{1,5}^2  x_{6,10}^{2}x_{3,8}^{2}\:\: \dif^4 x_{0}\, \dif^4x_{0'} }{x_{1,0}^{2} x_{3,0}^{2} x_{5,0}^{2}x_{0,0'}^{2}x_{6,0'}^{2}x_{8,0'}^{2}x_{10,0'}^{2}}
 \:, 
\end{align}
which is finite and depends on 7 cross-ratios
\begin{gather} 
    u_{1}=\frac{x_{3,6}^{2}x_{8,10}^{2}}{x_{3,8}^{2}x_{6,10}^{2}}\:, \qquad 
    u_{2}=\frac{x_{1,3}^{2}x_{5,8}^{2}}{x_{1,5}^{2}x_{3,8}^{2}} \:, \qquad 
    v_{1}=\frac{x_{6,8}^{2}x_{3,10}^{2}}{x_{3,8}^{2}x_{6,10}^{2}} \:, \qquad 
    v_{2}=\frac{x_{3,5}^{2}x_{1,8}^{2}}{x_{1,3}^{2}x_{5,8}^{2}} \:, \qquad  \nonumber \\
    u_{3}=\frac{x_{1,6}^{2}x_{8,10}^{2}}{x_{1,8}^{2}x_{6,10}^{2}} \:, \qquad 
    u_{4}=\frac{x_{6,8}^{2}x_{5,10}^{2}}{x_{5,8}^{2}x_{6,10}^{2}} \:, \qquad 
    u_{5}= \frac{x_{1,5}^{2}x_{6,10}^{2}}{x_{1,6}^{2}x_{5,10}^{2}} \:.  \label{uvdef}
\end{gather}
One can easily see that this integral reduces to $x_{1,5}^{2}x_{3,7}^{2} I_{\text{db}}$ in the collinear limit $x_{10}\to x_{8}$, $x_{8}\to x_{7}$ at the integrand level. The elliptic double-box integral has been evaluated as elliptic multiple polylogarithms in~\cite{Kristensson:2021ani} where its symbol is also given. However, a better start point turns out to be its one-fold integral representation given in \cite{Kristensson:2021ani}
\begin{equation} \label{elldb-onefoldint}
    I_{\text{ell-db}}= \int_{0}^{\infty} \frac{\dif x}{y} \mathcal{G}_{3}(x,y) \:,
\end{equation}
where $\mathcal{G}_{3}(x,y)$ is a known MPL of weight 3, which depends on 7 cross-rations, as well as $\{x,y\}$ given by the elliptic curve 
\begin{align}
    y^{2}
=\biggl(\frac{v_{1}}{u_{4}}\bigl((1{-}u_{4})(x{+}1{-}v_{2}){-}u_{1}{+}u_{3}v_{2}\bigr){+}h_{1}{+}h_{2}\biggr)^{2}{-}4h_{1}h_{2}, \label{ellcurve}
\end{align}
with
\begin{equation}
 \begin{aligned}
    h_{1}&=\frac{u_{2}u_{4}}{v_{1}}\bigl(x^{2}+(1{-}u_{1}{+}v_{1})x+v_{1}\bigr),  \\
    h_{2}&=\Bigl(x{+}\frac{v_{1}}{u_{4}}\Bigr)\Bigl((1{+}x{-}u_{1})\Bigl(\frac{u_{2}u_{4}}{v_{1}}-1\Bigr)+(1{-}u_{3})v_{2}\Bigr). 
\end{aligned}
\end{equation}

Now let us turn to the 8-point double box.
With the DCI regulator eq.\eqref{DCI-regulator}, the 7 cross-ratios in eq.\eqref{uvdef} behave like 
\begin{gather}
    u_{1}\to \epsilon \frac{x_{1,7}^{2}x_{3,6}^{2}}{x_{1,6}^{2}x_{3,7}^{2}}\:, \quad 
    u_{2} \to \frac{x_{1,3}^{2}x_{5,7}^{2}}{x_{1,5}^{2}x_{3,7}^{2}} =u\:, \quad 
    v_{1} \to \epsilon \frac{x_{3,8}^{2}x_{5,7}^{2}}{x_{3,7}^{2}x_{5,8}^{2}} \:, \quad 
    v_{2} \to \frac{x_{3,5}^{2}x_{1,7}^{2}}{x_{1,5}^{2}x_{3,7}^{2}} =v \:, \nonumber \\
    u_{3}\to \epsilon \:, \quad u_{4} \to \epsilon \:, \quad 
    u_{5} \to \frac{x_{1,5}^{2}x_{6,8}^{2}}{x_{1,6}^{2}x_{5,8}^{2}} \:, \label{uv-collinear}
\end{gather}
as $\epsilon\to0$. Correspondingly, the elliptic curve eq.\eqref{ellcurve} degenerates to
\begin{equation}
    y^{2}=\Bigl(x^{2}+x(1+u-v)x+u\Bigr)^{2}= \Bigl((x+z)(x+\bar{z})\Bigr)^{2}\:,
\end{equation}
and $\mathcal{G}_{3}(x,y)$ in eq.\eqref{elldb-onefoldint} simplifies dramatically as
\begin{equation}
\mathcal{G}_{3}(x,y) = \log^{2}(\epsilon) \log \frac{(1+x)(x+u)}{xv} + \log(\epsilon) \: \mathcal{G}_{2}^{\text{fin}} + \mathcal{G}_{3}^{\text{fin}} + O(\epsilon)\:,
\end{equation}
where $\mathcal{G}_{2,3}^{\text{fin}}$ are MPLs of weight 2 and 3, respectively.

A nice observation is that for the $\log^2 \epsilon$ term, the integral $\int \log \frac{(1+x)(x+u)}{xv} \frac{\dif x}{(x+z)(x+\bar{z})}$ indeed evaluates to the four-mass box integral $\frac{1}{\Delta}\widehat{\mathbf{A}}_{8,2}^{(1)}$ in eq.\eqref{boxInt}. This already provides some preliminary evidence that IR divergences will be cancelled in \eqref{ratio_function_component} to give a finite result. It is straightforward to evaluate the remaining integrals $\int \mathcal{G}_{2,3}^{\text{fin}}\frac{\dif x}{(x+z)(x+\bar{z})}$ with, say \texttt{HyperInt} or \texttt{PolyLogTools}~\cite{Panzer:2014caa,Duhr:2019tlz}. Here we only record the \emph{symbol}~\cite{Goncharov:2010jf} of its cyclic image $\widehat{I}_{\text{db}}(x_5,x_7,x_1,x_2,x_3,x_4)$, which can be obtained much easier by the algorithm provided in the Appendix
A of \cite{Caron-Huot:2011dec}, in the ancillary file \texttt{n82Lnnmhv_scale_component.txt} as variables \texttt{puredb\$log2e}, \texttt{puredb\$loge} and \texttt{puredb\$finite}.

\subsection*{The penta-box integral}
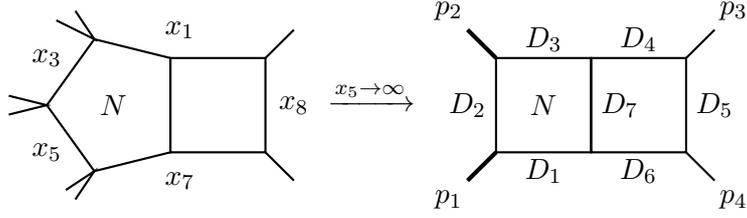
\begin{figure} 
\begin{center}
\begin{tikzpicture}[baseline={([yshift=-.5ex]current bounding box.center)},scale=0.25]
    \draw[black,thick] (-9,6)--(-5,5) -- (-5,0) --(-9,-1)--(-11.5,2.5)--cycle;
    \draw[black,thick] (-10,7.5)--(-9,6)--(-11,7);
    \draw[black,thick] (-10,-2.5)--(-9,-1)--(-10.5,-2);
    \draw[black,thick] (-5,0)--(-5,5)--(0,5)--(0,0)--cycle;
    \draw[black,thick] (1.5,6.5)--(0,5);
    \draw[black,thick] (1.5,-1.5)--(0,0);
    \draw[black,thick] (-13.5,3)--(-11.5,2.5)--(-13.5,2);
    \node at (-4.5,-1.5) {$x_7$};
    \node at (1.5,2.5) {$x_8$};
    \node at (-4.5,6.5) {$x_1$};
    \node at (-11.5,5) {$x_3$};
    \node at (-11.5,0) {$x_5$};
\node at (-8,2.5) {$N$};
\end{tikzpicture}
$\xrightarrow{x_{5}\to\infty}$
\begin{tikzpicture}[baseline={([yshift=-.5ex]current bounding box.center)},scale=0.25]
    \draw[black,thick] (0,5)--(-5,5)--(-5,0)--(0,0)--cycle;
    \draw[black,ultra thick] (-11.5,6.5)--(-10,5);
    \draw[black,ultra thick] (-11.5,-1.5)--(-10,0);
    \draw[black,thick] (-5,0)--(-5,5)--(-10,5)--(-10,0)--cycle;
    \draw[black,thick] (0,5)--(1.5,6.5);
    \draw[black,thick] (1.5,-1.5)--(0,0);
\node at (-12.5,-2.5) {$p_1$};
\node at (-12.5,7.5) {$p_2$};
\node at (2.5,7.5) {$p_3$};
\node at (2.5,-2.5) {$p_4$};
\node at (-7.5,2.5) {$N$};
\node at (-7.5,-1) {$D_1$};
\node at (-11.5,2.5) {$D_2$};
\node at (-7.5,6) {$D_3$};
\node at (-2.5,6) {$D_4$};
\node at (1.5,2.5) {$D_5$};
\node at (-2.5,-1) {$D_6$};
\node at (-3.5,2.5) {$D_7$};
\end{tikzpicture}
\end{center}
\caption{The penta-box integral of interest and the corresponding non-DCI double-box integral. Here $N$ represents a factor $x_{0,8}^{2}$ in the numerator of both integrals.} \label{pbandnDCIdb}
\end{figure}

Since $I_{\rm pb}$ can be obtained as double-collinear limits of $n=10$ penta-box integrals, in principle we can compute the latter similarly in (DCI) Feynman-parameterized form, but the computation turns out to be rather tedious. Fortunately, we find that they belong to the family of integrals recently computed in~\cite{He:2022ctv} by canonical differential equations: the DCI penta-box integral can be identified as a non-DCI double box integral by sending $x_{5}$ to $\infty$, see figure \ref{pbandnDCIdb}. This so-called non-DCI limit gives us a bijection between the kinematics of these two Feynman integrals,
\begin{equation}\label{bijection_kinematics}
m_1^2 \leftrightarrow x_{3,7}^2,\, m_2^2 \leftrightarrow x_{1,3}^2,\, m_3^2 \leftrightarrow x_{8,1}^2,\, m_4^2 \leftrightarrow x_{7,8}^2,\, s\leftrightarrow x_{1,7}^2,\, t\leftrightarrow x_{3,8}^2,\,N=x_{0,8}^2,
\end{equation}
and for external Mandelstam variables, namely $x_{i,j}^2$ with $i,j\neq 0,0'$, we further write them in terms of momentum-twistor variables
\begin{equation}
x_{i,j}^2=\frac{\langle i{-}1\ i\ j{-}1\ j\rangle}{\langle i{-}1\ i\ I_{\infty}\rangle \langle j{-}1\ j\ I_{\infty}\rangle}\leftrightarrow \frac{\langle i{-}1\ i\ j{-}1\ j\rangle}{\langle i{-}1\ i\ 4\ 5\rangle \langle j{-}1\ j\ 4\ 5\rangle},
\end{equation}
where the infinite bi-twistor $I_{\infty}$ is given by $(45)$ since we have taken $x_5\to \infty$.

The non-DCI double-box integral $I_{\text{nDCI-db}}$ is denoted in \cite{He:2022ctv} as $G_{1,1,1,1,1,1,1,-1}$, which can be expressed as linear combination of the following three master integrals 
\begin{align*}
&I_5= r_9 G_{0,1,1,1,1,1,1,0},\quad I_6= r_8 G_{1,1,0,1,1,1,1,0}\\
&I_3= r_2 (-m_3^2 G_{0,1,1,1,1,1,1}-m_4^2 G_{1,1,0,1,1,1,1}+s G_{1,1,1,1,1,1,1,-1}),
\end{align*}
where we have used (see the figure \ref{pbandnDCIdb} above for definition of $D_i$'s):
\[
    G_{\alpha_1,\dots,\alpha_7,\alpha_{8}}:=\int \frac{{\rm d}^D \ell_1 {\rm d}^D \ell_2 }{(i \pi^{D/2})^{2}}\frac{1}{D_1^{\alpha_1}\cdots D_7^{\alpha_7}N^{\alpha_{8}}}.
\]
Therefore, $I_{\text{nDCI-db}}=G_{1,1,1,1,1,1,1,-1}$ can be solved as 
\begin{equation}\label{ndcidb}
I_{\text{nDCI-db}}=\frac{1}{r_2 s}I_3+\frac{m_3^2}{r_9 s}I_5+\frac{m_4^2}{r_8 s}I_6,
\end{equation}
where $I_3$, $I_5$ and $I_6$ are three pure weight-4 MPL functions, and $r_2,r_8,r_9$ are three different square roots of kinematics variables given by
\begin{align*}
&r_2^2=(s-m_1^2-m_2^2)^2-4 m_1^2 m_2^2,\\
&r_8^2=((m_1^2-m_2^2) m_4^2+ s (t-m_1^2))^2-4 s m_1^2 m_3^2 m_4^2,\\
&r_9^2=((m_2^2-m_1^2) m_3^2+ s (t-m_2^2))^2-4 s m_2^2 m_3^2 m_4^2.
\end{align*}
Applying the bijection eq.\eqref{bijection_kinematics} and then taking double-collinear limits with DCI regulator leads to
\begin{equation}\label{col_lim}
m_3^2\leftrightarrow x_{8,1}^2=\frac{x_{7,1}^2 x_{8,2}^2}{x_{7,2}^2}\epsilon,\quad 
m_4^2\leftrightarrow x_{7,8}^2=\frac{x_{6,8}^2 x_{7,1}^2}{x_{6,1}^2}\epsilon,
\end{equation}
we then get the full result of penta-box integral $I_{\text{pb}}$ in the DCI regularization. 

Under the collinear limits eq.\eqref{col_lim}, only the square root $r_2$ survives:
\begin{equation} 
r_2 \to x_{3,7}^2 \Delta,\quad r_8\to x_{1,7}^2(x_{3,8}^2-x_{3,7}^2),\quad r_9\to x_{1,7}^2(x_{3,8}^2-x_{1,3}^2),
\end{equation}
and this implies that the divergent penta-box integral shares the same overall factor, the square root $(x_{1,5}^{2}x_{3,7}^{2}\Delta)^{-1}$, with the divergent DCI double-box integral. Since the coefficients of the last two terms in eq.\eqref{ndcidb} vanish, we here only need to consider the collinear limit of $I_3$. By taking the collinear limit of its letters and keeping the leading terms of $\epsilon$ in the symbol, we find that $\widehat{I}_{\text{pb}}$ has the  expected IR divergence structure and takes the form
\begin{equation}
\widehat{I}_{\text{pb}}=\widehat{\mathbf{A}}_{8,2}^{(1)}\log^2(\epsilon)+I_{\text{pb}}^{(-1)}\log(\epsilon)+I_{\text{pb}}^{(0)},
\end{equation}
where $I_{\text{pb}}^{(-1)}$ and $I_{\text{pb}}^{(0)}$ are two finite, pure MPL functions of weight $3$ and $4$ respectively. In the ancillary file \texttt{n82Lnnmhv_scale_component.txt}, the three terms of 
$\mathcal{S}(\widehat{I}_{\text{pb}}(x_3,x_5,x_7,x_1,x_2))$ are stored in three variables \texttt{purepb\$log2e}, \texttt{purepb\$loge} and \texttt{purepb\$finite}.

\subsection*{The subtraction}

Now we move to the subtraction in \eqref{ratio_function_component} which should give an IR finite ratio function. The cancellation of leading divergences, {\it i.e.} the $\log^2 (\epsilon)$ part, is easy to see. Since the $\log^2(\epsilon)$ divergence of both penta-box and double-box integrals is given by the pure function $\widehat{\mathbf{A}}_{8,2}^{(1)}$, by summing over their cyclic images (see eq.\eqref{An0}) in the ratio function, we see that $\log^2(\epsilon)$ divergence is cancelled perfectly. 

What is much more non-trivial, is the cancellation of sub-leading $\log(\epsilon)$ divergences in the ratio function. Note that the $\log(\epsilon)$ part of both double-box and penta-box integrals is given by complicated weight-$3$ MPL functions, which certainly look much more complicated than the subtraction term in \eqref{ratio_function_component}. We find it totally remarkable that in the cyclic sum, all these complicated weight-3 MPL functions combine into precisely the $\log(\epsilon)$ part of the eq.\eqref{An0}, namely simple logarithmic functions times $\widehat{\mathbf{A}}_{8,2}^{(1)}$. This provides a very non-trivial consistency check of our computations for both DCI-regulated integrals, and such a precise cancellation of all IR divergences produces a finite result for the ratio function. We record its symbol in \texttt{scalecomponent} in the ancillary file \texttt{n82Lnnmhv_scale_component.txt}, and we also provide its  alphabet as the variable \texttt{alphabet} for convenience.

\section{The symbol and alphabet of the ratio function}

In this section, we discuss the symbol and alphabet of the ratio function, ${\widehat{\bf R}}^{(2)}_{8,2}$. The full symbol is too lengthy to be present here, and we record it in an ancillary file. Here we briefly comment on certain structures we find in the symbol of 8-point N${}^2$MHV, especially compared with those for NMHV/MHV cases. 

A new feature of this component of N${}^2$MHV amplitudes, as opposed to other components of NMHV/MHV amplitudes, is that the symbol is odd with respect to the flip $\Delta \to -\Delta$. This is due to the fact that its leading singularity is proportional to (the inverse of) $\Delta$, thus, similar to the one-loop component (four-mass box function) but for the first time for two-loop amplitudes, each term of the symbol (a ``word") contains {\it odd} number of {\it odd} letters, which are algebraic (non-rational) functions of momentum twistors. Recall that for one-loop case, the symbol reads
\begin{equation}\label{4msymbol}
{\cal S}[\widehat{I}_{\text{bl}}^{(1)}(x_1, x_3, x_5, x_7)]=u \otimes \frac{z}{\bar{z}} - v \otimes \frac{1-z}{1-\bar{z}}\,,
\end{equation}
which is odd under $\Delta \to -\Delta$ since it contains the odd letters $z/\bar{z}, (1-z)/(1-\bar{z})$ in the second entry, {\it e.g.} $\log (z/\bar{z}) \to - \log(z/\bar{z})$ under the flip. 

Similarly at two-loop level, since the first entry must be even letters, there is either one such odd letter at the second, third and fourth entry, or all these three entries are odd. In particular, this is the first example of ${\cal N}=4$ SYM amplitudes at two-loop level where the last entry can be odd letters. Recall that for the two-loop 8-point NMHV amplitude we find $9$ multiplicatively independent algebraic letters that are odd in $\Delta \to -\Delta$) associated with this square root (and $9$ more related by a cyclic rotation $i\to i{+}1$). It is nice that this component of the two-loop 8-point N${}^2$MHV amplitude still contains exactly this $9$-dimensional space of odd letters, and we denote them as 
\begin{equation}\label{odd_letters}
\{\chi(a_1),\chi(a_2)\}|_{i\to i+2k} \text{ for $k=0,1,2,3$ and } \frac{\chi(0)}{\chi(1)},
\end{equation}
where $\chi(a):=(z-a)/(\bar z-a)$ and 
\[
a_1=\frac{\langle 4567\rangle \langle 1258\rangle}{\langle 2567\rangle \langle 1458\rangle},\quad
a_2=\frac{\langle 4567\rangle \langle 1358\rangle}{\langle 3567\rangle \langle 1458\rangle}.
\]
Our result strongly supports the conjecture that altogether the $9+9$ letters constitute all algebraic (odd) letters for $n=8$ amplitudes for all helicity sectors. 

Next we move to even letters, which are rational functions of momentum twistors. We note that again almost all of them (together with those for the other component, related by $i \to i{+}1$) have been seen in the alphabet of $n=8$ NMHV amplitudes: they are $92$ polynomials of Pl\"{u}cker coordinates, which can form $92-8=84$ DCI combinations. There are $56$ Pl\"{u}cker coordinates ($4$-brackets): under cyclic rotations $i\to i{+}2$ they form $13$ length-$4$ orbits and $2$ length-$2$ ones, with $15$ seeds chosen as
\begin{align}
&\langle 1 2 3 4\rangle, \langle 1 2 3 5\rangle,\langle 1 2 3 6\rangle,\langle 1 2 3 7\rangle, \langle 1 2 3 8\rangle,\langle 1 2 4 5\rangle, \langle 1 2 4 8\rangle, \underline{\langle 1 2 5 6\rangle}, \langle 1 2 5 8\rangle,\langle 1 2 6 7\rangle, \langle 1 2 6 8\rangle, \\\nonumber
&\langle 1 3 5 8\rangle, \langle 1 3 6 8\rangle , \langle 1 4 6 8\rangle, \underline{\langle 1 4 5 8\rangle}\,,
\end{align}
where two brackets with underline generate length-$2$ orbits under $i \to i{+}2$ rotations. 
The remaining $36$ letters are quadratic in Pl\"{u}cker coordinates, which form $9$ length-$4$ orbits with seeds
\begin{align}
&\langle 1 (23) (45) (67)\rangle, \langle 1 (23) (45) (78)\rangle,   \langle 1 (28) (34) (67)\rangle,   \langle 1 (28) (45) (67)\rangle, \\\nonumber
&\langle 2 (13) (45) (67)\rangle, \langle 2 (13) (45) (78)\rangle,   \langle 2 (18) (34) (67)\rangle,   \langle 2 (18) (45) (67)\rangle, \langle 812 (67) \cap (345) \rangle\,,
\end{align}
where we have defined 
\begin{align}
&\langle i j k (ab) \cap (cde)\rangle:= \langle i j k a \rangle \langle b c d e\rangle-\langle i j k b \rangle \langle a c d e\rangle\\
&\langle i (ab)(cd)(ef)\rangle:=\langle i a c d\rangle \langle i b e f \rangle-\langle i a e f\rangle \langle i b c d \rangle.
\end{align}
In addition to these letters which have appeared in NMHV/MHV amplitudes, we find exactly one new letter, which is nothing but the inverse of leading singularity, $\Delta$!

Let us briefly comment on how the letters appear in the symbol. As indicated by physical discontinuity conditions~\cite{Gaiotto:2011dt}, we find exactly $20$ letters of the form $\langle i \,i{+}1\, j\, j{+}1 \rangle$ which appear in the first entry; the frozen ones $\langle i\, i{+}1\, i{+}2\, i{+}3\rangle$ (for $i=1, \cdots, 8$, treated as $2$ length-$4$ orbits above), and $3$ length-$4$ orbits generated by seeds $\langle 1245\rangle$, $\langle 1256\rangle$ and $\langle 1267\rangle$. The second entry respects Steinmann relations, and we find that all first two entries are consistent with the general prediction of~\cite{He:2021mme}: they are either $\log \log$ terms respecting Steinmann relations, or the symbol of one-loop box functions {\it i.e.}  $I^{(1)}(x_1, x_3, x_5, x_7)$ and finite part of its lower-mass degenerations
\[
\operatorname{Li}_{2}\biggl(1-\frac{x_{a,c}^{2}x_{a-1,b}^{2}}{x_{a-1,c}^{2}x_{a,b}^{2}}\biggr) \:,
\]
with $a+1<b<c<a-2$ in a cyclic order.
The third entry contains almost all letters in the alphabet, except for $4+4$ quadratic letters generated by $\langle 1 (28)(34)(67)\rangle$ and $\langle 2(13)(45)(78) \rangle$ which have appeared in the second entry. Finally, we find the following letters for the last entry: the $9$ algebraic (odd) letters of eq.\eqref{odd_letters}, $\Delta$, $4+4$ quadratic ones generated by $\langle 1 (23) (45) (67)\rangle$ and $\langle 2 (81) (45) (67)\rangle $, and the following $34$ Pl\"{u}cker coordinates: the frozen ones $\langle i \,i{+}1 \,i{+}2\,i{+}3\rangle$, $6$ length-$4$ orbits with seeds (all of the form $\langle \bar{i} j\rangle:=\langle i-1\,i\,i+1\,j \rangle$, originated from MHV amplitude in the subtraction term):
\begin{equation}
\langle 1235\rangle,  \langle 1236\rangle,  \langle 1237\rangle,   \langle 1248\rangle, \langle 1258\rangle,     \langle 1268\rangle\,. 
\end{equation}
and a length-$2$ orbit $\{\langle 1458\rangle$, $\langle 2367\rangle\}$.

We have observed more structures for terms in the symbol involving odd letters. 
In particular, those terms with $3$ odd letters are very simple: the first-two entries can only be the symbol of the four-mass box, eq.\eqref{4msymbol}, and we record the last-two (odd) entries: in principle we have $9\times 9$ terms but the result is much more compact, which we record as
\[
\sum_{i,j=1}^9 c_{ij}v_i\otimes v_j \:,
\]
where
\[
(c_{ij})=\frac{1}{4}\begin{pmatrix}
 6 & 1 & 0 & 1 & -4 & -2 & 0 & 2 & -2 \\
 1 & 6 & 1 & 0 & 2 & -4 & -2 & 0 & -2 \\
 0 & 1 & 6 & 1 & 0 & 2 & -4 & -2 & -2 \\
 1 & 0 & 1 & 6 & -2 & 0 & 2 & -4 & -2 \\
 -4 & 2 & 0 & -2 & 8 & -2 & 0 & -2 & 1 \\
 -2 & -4 & 2 & 0 & -2 & 8 & -2 & 0 & 1 \\
 0 & -2 & -4 & 2 & 0 & -2 & 8 & -2 & 1 \\
 2 & 0 & -2 & -4 & -2 & 0 & -2 & 8 & 1 \\
 -2 & -2 & -2 & -2 & 1 & 1 & 1 & 1 & 6 \\
\end{pmatrix}
\]
and
\[
\{v_1,\cdots,v_9\}=\biggl\{\chi(a_1),\chi(a_1)|_{i\to i+2},\cdots,\chi(a_2),\chi(a_2)|_{i\to i+2},\cdots,\frac{\chi(0)}{\chi(1)}\biggr\}.
\]
Alternatively, if we denote $\chi_i:=v_i$ and $\chi'_i:=v_{i{+}4}$ for $i=1,\cdots, 4$ and $\chi'':=v_9$, we can write this symmetric, weight-$2$ word as the symbol of
\begin{align}
G=&\frac 1 4 \sum_{i=1}^4 \left(3\log^2 \chi_i +\log\chi_i \log \chi_{i{+}1}+4 \log^2\chi'_i-2\log \chi'_i \log\chi'_{i{+}1}+3 \log^2\chi''-\log\chi'' \log\biggl(\frac{\chi_i^2}{\chi'_i}\biggr)\right)\nonumber \\
&-\sum_{i=1}^4 \left(\log\chi_i \log \chi'_i+\frac 1 2 \log \chi_i \log \frac{\chi'_{i{+}1}}{\chi'_{i{-}1}}\right)\,,
\end{align}
where the symmetry under the cyclic rotation $i\to i+2$ is manifest. The part of the complete symbol which contains $3$ odd letters can be put in a remarkably compact form: 
\begin{equation}\label{3odd}
{\cal S}[\widehat{{\bf R}}^{(2)}_{8,2}]={\cal S}[\widehat{I}_{\text{bl}}^{(1)}] \otimes {\cal S}[G] + \text{terms with one odd letter}\,,   
\end{equation}
where the remaining terms have one odd letter in either second, third or last entry. These terms seem to be significantly more complicated. For example, for terms where only the second entry is odd ($\chi(0)$ or $\chi(1)$), the first two entries must again form the symbol of $\widehat{I}_{\text{bl}}^{(1)}$, while the last two entries consist of even letters, including  the new letter $\Delta$.

\section{Conclusion and Outlook}
In this paper, we have computed a particularly nice component of two-loop N${}^2$MHV amplitudes, whose leading singularity is proportional to the inverse of the four-mass square root, which makes it distinct from other N${}^2$MHV components. The component receives contributions from only two new integrals, which we have computed using the DCI regularization and the finite ratio function is obtained by subtracting IR divergences. This is the first two-loop amplitudes beyond MHV and NMHV cases that have been computed, and the resulting symbol has several interesting features; compared to NMHV case it contains exactly one new letter, which is \emph{the square root itself}, and either one or all of the last three entries must be algebraic (odd) letters. We have found a remarkable simplicity for the latter case. 

There are several directions worth further investigations. Two immediate generalizations are (1) computing all components of the two-loop 8-point N${}^2$MHV amplitude, which should be straightforward but rather tedious; (2) computing such special components for $n$-point N${}^2$MHV amplitudes, {\it e.g.} $9$ such components each with a square root for $n=9$ case; our techniques should be applicable but several new cases of penta-box integrals {\it etc.} are needed. It would be interesting to look for more structures in the symbol and alphabet of such special components for $n=8$ and higher, similar to what we have considered for NMHV and MHV cases; it would be nice to further study mathematical structures in the symbol of our two-loop integrals (see ~\cite{He:2021eec, He:2021esx, He:2021mme, Yang:2022gko, He:2022tph} and references therein). Another important question is how to uplift these symbols to functions and directly bootstrap them.  Now that we have some control of $n=8$ MHV, NMHV and these components of N${}^2$MHV amplitudes, it would also be intriguing to compare them in more details. 

Relatively, it would be highly desirable to study what do such structures, especially the appearance of algebraic letters in the last entries, mean for the ${\bar Q}$ and $Q^{(1)}$ anomaly equations, and eventually use them for computing amplitudes with $k\geq 2$. In this regard, it would be nice to apply these anomaly equations for certain components rather than for the complete super-amplitude (related anomaly equations have been applied to other theories and even individual integrals in ~\cite{Chicherin:2022gky,Corcoran:2021gda}). We have also demonstrated how DCI-regulated integrals are useful for computing IR-safe quantities for loop amplitudes with higher multiplicities. It would be interesting to push this direction further, including the computation of non-planar SYM amplitudes and those in supergravity theories~\cite{Bourjaily:2019iqr,Bourjaily:2019gqu}, as well as amplitudes in other theories (such as ABJM, see~\cite{Caron-Huot:2012sos,He:2022toappear}). 

\acknowledgments

We thank Rourou Ma, Zihao Wu, Qinglin Yang and Yang Zhang for collaborations on related projects. SH is supported by National Natural Science Foundation of China under Grant No. 11935013, 11947301, 12047502, 12047503 and the Key Research Program of the Chinese Academy of Sciences, Grant NO. XDPB15. CZ was supported in part by the ERC starting grant 757978 and grant 00025445 from the Villum Fonden.

\appendix

\section{Review of multiple polylogarithm and its symbol}

Many amplitudes and Feynman integrals in QFT can be described by multiple polylogarithms. In this appendix, we briefly review some basic facts about multiple polylogarithms.

Multiple polylogarithms are generalizations of classical polylogarithms, such as $\log$, $\operatorname{Li}_2$, $\operatorname{Li}_3$, $\dots$, which are defined by~\cite{goncharov2005galois}
\begin{equation}
    G(a_{1},\ldots,a_{n};z):=\int_{0}^{z}\dif \log(t-a_{1})\, G(a_{2},\ldots,a_{n};t), \label{Gpolylot} 
\end{equation} 
with $G(;z):=1$ and exceptional cases
\[
    G(\underbrace{0, \ldots, 0}_{k} ; z):=\frac{1}{k !}(\log z)^{k},
\]
where $n$ is called the \textit{weight} of $G(a_{1},\ldots,a_{n};z)$. For example,
\[
    G(0;x)=\log x,\quad G(1;x)=\log(1-x),\quad G(0,1;x)=-\operatorname{Li}_2(x),\quad \dots.
\]
It is straightforward to see that the total differential of eq.\eqref{Gpolylot} satisfies
\begin{equation}
    \dif G(a_1,\dots,a_n;z) = \sum_{i=1}^n G(a_1,\dots,\hat{a}_{i},\dots,a_n;z)\,\dif \log \frac{a_i-a_{i{-}1}}{a_i-a_{i{+}1}},
\end{equation}
where $a_i$ is deleted in the $i$-th summand with boundary cases $a_0:=z$ and $a_{n{+}1}:=0$. 

There is a well-known Hopf algebra structure~\cite{Goncharov:2005sla} on multiple polylogarithms, which has led to the notion of \textit{symbol}~\cite{Goncharov:2010jf, Duhr:2011zq}. For any weight-$n$ multiple polylogarithm $G^{(n)}$ whose differential reads
\[
\dif G^{(n)}=\sum_i G^{(n-1)}_i \dif\log x_i,
\]
where $\{G^{(n-1)}_i\}$ are polylogarithms of lower weight $(n{-}1)$, its symbol $\mathcal{S}(G^{(n)})$ is recursively defined by
\[
\mathcal{S}(G^{(n)}):=\sum_i \mathcal{S}(G^{(n-1)}_i)\otimes x_i.
\]
For example, 
\[
\mathcal{S}(\log(x))=x,\quad 
\mathcal{S}(\operatorname{Li}_2(x))=-\,(1-x)\otimes x.
\]
The entries of a symbol are called its \textit{letters}, and the collection of all letters is called its \textit{alphabet}.

For our calculation, it's more convenient to perform integrations on the symbol level directly based on the following rules~\cite{CaronHuot:2011kk}: Suppose we have an integral
\[
\int_a^b {\rm d}\log(t+c)\, (F(t)\otimes w(t)),
\]
where $F(t)\otimes w(t)$ is a linear reducible symbol in $t$, {\it i.e.} its entries are products of powers of linear polynomials in $t$, and $w(t)$ is the last entry. The total differential of this integral is the sum of the following two parts:
\begin{compactenum}[\quad (1)]
\item the contribution from endpoints:
\begin{equation}
    {\rm d}\log(t+c)\Bigl(F(t)\otimes w(t)\Bigr)\Bigr|_{t=a}^{t=b} \Rightarrow \Bigl(F(t)\otimes w(t)\otimes (t+c)\Bigl)\Bigr|_{t=a}^{t=b}, \label{symbolint1}
\end{equation}
\item contributions from the last entry: for a term where $w(t)$ is a constant,
\begin{equation}
\left(\int_a^b {\rm d}\log(t+c)\, F(t)\right){\rm d}\log w \Rightarrow\left(\int_a^b {\rm d}\log(t+c)\, F(t)\right)\otimes w, \label{symbolint2}
\end{equation}
and for a term where $w(t)=t+d$,
\begin{equation}
\left(\int_a^b {\rm d}\log \frac{t+c}{t+d}\, F(t)\right){\rm d}\log (c-d)
 \Rightarrow\left(\int_a^b {\rm d}\log \frac{t+c}{t+d}\, F(t)\right)\otimes (c-d). \label{symbolint3}
\end{equation}
\end{compactenum}

\section{Alphabet of 3-loop MHV octagon and 2-loop NMHV octagon}

The symbol alphabet for the three-loop MHV BDS-normalized octagon   $(\mathcal{A}_{8,0}/\mathcal{A}_{8}^{\text{BDS}})^{(3)}$ consists of $204$ multiplicative-independent rational letters and 18 independent DCI algebraic letters \cite{Li:2021bwg}. The 204 rational letters are organized as follows
    \begin{itemize}
        \item $\binom{8}{4}-2=68$ : all $\langle abcd\rangle$ except $\langle 1357\rangle$ and $\langle 2468\rangle$;
        \item 1 cyclic class of $\langle 12(345)\cap (678)\rangle$;
        \item 7 cyclic classes of $\langle 1(ij)(kl)(mn)\rangle$ with $2\leq i<j<k<l<m<n\leq 8$;\\
         5 cyclic classes of $\langle 1(28)(kl)(mn)\rangle$ with $2<k<l<m<n< 8$;
        \item 5 cyclic classes of 
        $
        \langle \bar 2\cap \bar 4\cap (568) \cap \bar 8\rangle,
        \langle \bar 2\cap \bar 4\cap \bar 6 \cap (681)\rangle,
        \langle (127)\cap (235)\cap \bar 5 \cap \bar 7\rangle,
        \langle (127)\cap \bar 3\cap (356) \cap \bar 7\rangle,
        \langle \bar 2\cap (278)\cap (346) \cap \bar 6\rangle.
        $
    \end{itemize}
Here we introduce the notations $\bar a=(a{-}1,a,a{+}1)$ and
\begin{align*}
&\langle ab(cde)\cap (fgh)\rangle:=\langle abde\rangle \langle cfgh\rangle+\langle abec\rangle \langle dfgh\rangle+\langle abcd\rangle \langle efgh\rangle,\\
&\langle (a_1b_1c_1)\cap (a_2b_2c_2)\cap (a_3b_3c_3)\cap (a_4b_4c_4)\rangle:=\langle (a_1b_1c_1)\cap (a_2b_2c_2),(a_3b_3c_3)\cap (a_4b_4c_4)\rangle.
\end{align*}
For the two-loop NMHV BDS-normalized octagon~\cite{Zhang:2019vnm}, there are only 180 rational letters, the following 24 rational letters do not appear,
\[\text{cyclic images of } \langle 1(23)(46)(78)\rangle,\: \langle \bar 2\cap \bar 4\cap (568) \cap \bar 8\rangle \text{ and } \langle \bar 2\cap \bar 4\cap \bar 6 \cap (681)\rangle.
\]
The algebraic letters are the same for the 3-loop MHV octagon and the two-loop NMHV octagon, which are cyclic images of eq.\eqref{odd_letters} by taking $i\to i+1$.

\bibliographystyle{JHEP}
\bibliography{bibtex}

\providecommand{\href}[2]{#2}\begingroup\raggedright\begin{thebibliography}{10}

\bibitem{ArkaniHamed:2010kv}
N.~Arkani-Hamed, J.~L. Bourjaily, F.~Cachazo, S.~Caron-Huot, and J.~Trnka, {\it
  {The All-Loop Integrand For Scattering Amplitudes in Planar N=4 SYM}},  {\em
  JHEP} {\bf 01} (2011) 041, [\href{http://arxiv.org/abs/1008.2958}{{\tt
  arXiv:1008.2958}}].

\bibitem{Arkani-Hamed:2012zlh}
N.~Arkani-Hamed, J.~L. Bourjaily, F.~Cachazo, A.~B. Goncharov, A.~Postnikov,
  and J.~Trnka, {\em {Grassmannian Geometry of Scattering Amplitudes}}.
\newblock Cambridge University Press, 4, 2016.

\bibitem{Arkani-Hamed:2013jha}
N.~Arkani-Hamed and J.~Trnka, {\it {The Amplituhedron}},  {\em JHEP} {\bf 10}
  (2014) 030, [\href{http://arxiv.org/abs/1312.2007}{{\tt arXiv:1312.2007}}].

\bibitem{Dixon:2011pw}
L.~J. Dixon, J.~M. Drummond, and J.~M. Henn, {\it {Bootstrapping the three-loop
  hexagon}},  {\em JHEP} {\bf 11} (2011) 023,
  [\href{http://arxiv.org/abs/1108.4461}{{\tt arXiv:1108.4461}}].

\bibitem{Dixon:2014xca}
L.~J. Dixon, J.~M. Drummond, C.~Duhr, M.~von Hippel, and J.~Pennington, {\it
  {Bootstrapping six-gluon scattering in planar N=4 super-Yang-Mills theory}},
  {\em PoS} {\bf LL2014} (2014) 077,
  [\href{http://arxiv.org/abs/1407.4724}{{\tt arXiv:1407.4724}}].

\bibitem{Dixon:2014iba}
L.~J. Dixon and M.~von Hippel, {\it {Bootstrapping an NMHV amplitude through
  three loops}},  {\em JHEP} {\bf 10} (2014) 065,
  [\href{http://arxiv.org/abs/1408.1505}{{\tt arXiv:1408.1505}}].

\bibitem{Drummond:2014ffa}
J.~M. Drummond, G.~Papathanasiou, and M.~Spradlin, {\it {A Symbol of
  Uniqueness: The Cluster Bootstrap for the 3-Loop MHV Heptagon}},  {\em JHEP}
  {\bf 03} (2015) 072, [\href{http://arxiv.org/abs/1412.3763}{{\tt
  arXiv:1412.3763}}].

\bibitem{Dixon:2015iva}
L.~J. Dixon, M.~von Hippel, and A.~J. McLeod, {\it {The four-loop six-gluon
  NMHV ratio function}},  {\em JHEP} {\bf 01} (2016) 053,
  [\href{http://arxiv.org/abs/1509.08127}{{\tt arXiv:1509.08127}}].

\bibitem{Caron-Huot:2016owq}
S.~Caron-Huot, L.~J. Dixon, A.~McLeod, and M.~von Hippel, {\it {Bootstrapping a
  Five-Loop Amplitude Using Steinmann Relations}},  {\em Phys. Rev. Lett.} {\bf
  117} (2016), no.~24 241601, [\href{http://arxiv.org/abs/1609.00669}{{\tt
  arXiv:1609.00669}}].

\bibitem{Dixon:2016nkn}
L.~J. Dixon, J.~Drummond, T.~Harrington, A.~J. McLeod, G.~Papathanasiou, and
  M.~Spradlin, {\it {Heptagons from the Steinmann Cluster Bootstrap}},  {\em
  JHEP} {\bf 02} (2017) 137, [\href{http://arxiv.org/abs/1612.08976}{{\tt
  arXiv:1612.08976}}].

\bibitem{Drummond:2018caf}
J.~Drummond, J.~Foster, {\"{O}}.~G{\"{u}}rdo{\u{g}}an, and G.~Papathanasiou,
  {\it {Cluster adjacency and the four-loop NMHV heptagon}},  {\em JHEP} {\bf
  03} (2019) 087, [\href{http://arxiv.org/abs/1812.04640}{{\tt
  arXiv:1812.04640}}].

\bibitem{Caron-Huot:2019vjl}
S.~Caron-Huot, L.~J. Dixon, F.~Dulat, M.~von Hippel, A.~J. McLeod, and
  G.~Papathanasiou, {\it {Six-Gluon amplitudes in planar $ \mathcal{N} $ = 4
  super-Yang-Mills theory at six and seven loops}},  {\em JHEP} {\bf 08} (2019)
  016, [\href{http://arxiv.org/abs/1903.10890}{{\tt arXiv:1903.10890}}].

\bibitem{Caron-Huot:2019bsq}
S.~Caron-Huot, L.~J. Dixon, F.~Dulat, M.~von Hippel, A.~J. McLeod, and
  G.~Papathanasiou, {\it {The Cosmic Galois Group and Extended Steinmann
  Relations for Planar $\mathcal{N} = 4$ SYM Amplitudes}},  {\em JHEP} {\bf 09}
  (2019) 061, [\href{http://arxiv.org/abs/1906.07116}{{\tt arXiv:1906.07116}}].

\bibitem{Dixon:2020cnr}
L.~J. Dixon and Y.-T. Liu, {\it {Lifting Heptagon Symbols to Functions}},  {\em
  JHEP} {\bf 10} (2020) 031, [\href{http://arxiv.org/abs/2007.12966}{{\tt
  arXiv:2007.12966}}].

\bibitem{Anastasiou:2009kna}
C.~Anastasiou, A.~Brandhuber, P.~Heslop, V.~V. Khoze, B.~Spence, and
  G.~Travaglini, {\it {Two-Loop Polygon Wilson Loops in N=4 SYM}},  {\em JHEP}
  {\bf 05} (2009) 115, [\href{http://arxiv.org/abs/0902.2245}{{\tt
  arXiv:0902.2245}}].

\bibitem{CaronHuot:2011ky}
S.~Caron-Huot, {\it {Superconformal symmetry and two-loop amplitudes in planar
  N=4 super Yang-Mills}},  {\em JHEP} {\bf 12} (2011) 066,
  [\href{http://arxiv.org/abs/1105.5606}{{\tt arXiv:1105.5606}}].

\bibitem{Zhang:2019vnm}
S.~He, Z.~Li, and C.~Zhang, {\it {Two-loop Octagons, Algebraic Letters and
  $\bar{Q}$ Equations}},  {\em Phys. Rev. D} {\bf 101} (2020), no.~6 061701,
  [\href{http://arxiv.org/abs/1911.01290}{{\tt arXiv:1911.01290}}].

\bibitem{He:2020vob}
S.~He, Z.~Li, and C.~Zhang, {\it {The symbol and alphabet of two-loop NMHV
  amplitudes from $\bar{Q}$ equations}},  {\em JHEP} {\bf 03} (2021) 278,
  [\href{http://arxiv.org/abs/2009.11471}{{\tt arXiv:2009.11471}}].

\bibitem{Alday:2007hr}
L.~F. Alday and J.~M. Maldacena, {\it {Gluon scattering amplitudes at strong
  coupling}},  {\em JHEP} {\bf 06} (2007) 064,
  [\href{http://arxiv.org/abs/0705.0303}{{\tt arXiv:0705.0303}}].

\bibitem{Alday:2007he}
L.~F. Alday and J.~Maldacena, {\it {Comments on gluon scattering amplitudes via
  AdS/CFT}},  {\em JHEP} {\bf 11} (2007) 068,
  [\href{http://arxiv.org/abs/0710.1060}{{\tt arXiv:0710.1060}}].

\bibitem{Alday:2009yn}
L.~F. Alday and J.~Maldacena, {\it {Null polygonal Wilson loops and minimal
  surfaces in Anti-de-Sitter space}},  {\em JHEP} {\bf 11} (2009) 082,
  [\href{http://arxiv.org/abs/0904.0663}{{\tt arXiv:0904.0663}}].

\bibitem{Brandhuber:2007yx}
A.~Brandhuber, P.~Heslop, and G.~Travaglini, {\it {MHV amplitudes in N=4 super
  Yang-Mills and Wilson loops}},  {\em Nucl. Phys. B} {\bf 794} (2008)
  231--243, [\href{http://arxiv.org/abs/0707.1153}{{\tt arXiv:0707.1153}}].

\bibitem{Drummond:2007aua}
J.~M. Drummond, G.~P. Korchemsky, and E.~Sokatchev, {\it {Conformal properties
  of four-gluon planar amplitudes and Wilson loops}},  {\em Nucl. Phys. B} {\bf
  795} (2008) 385--408, [\href{http://arxiv.org/abs/0707.0243}{{\tt
  arXiv:0707.0243}}].

\bibitem{Drummond:2007cf}
J.~Drummond, J.~Henn, G.~Korchemsky, and E.~Sokatchev, {\it {On planar gluon
  amplitudes/Wilson loops duality}},  {\em Nucl. Phys. B} {\bf 795} (2008)
  52--68, [\href{http://arxiv.org/abs/0709.2368}{{\tt arXiv:0709.2368}}].

\bibitem{Drummond:2007bm}
J.~Drummond, J.~Henn, G.~Korchemsky, and E.~Sokatchev, {\it {The hexagon Wilson
  loop and the BDS ansatz for the six-gluon amplitude}},  {\em Phys. Lett. B}
  {\bf 662} (2008) 456--460, [\href{http://arxiv.org/abs/0712.4138}{{\tt
  arXiv:0712.4138}}].

\bibitem{Drummond:2008aq}
J.~Drummond, J.~Henn, G.~Korchemsky, and E.~Sokatchev, {\it {Hexagon Wilson
  loop = six-gluon MHV amplitude}},  {\em Nucl. Phys. B} {\bf 815} (2009)
  142--173, [\href{http://arxiv.org/abs/0803.1466}{{\tt arXiv:0803.1466}}].

\bibitem{Bern:2008ap}
Z.~Bern, L.~Dixon, D.~Kosower, R.~Roiban, M.~Spradlin, C.~Vergu, and
  A.~Volovich, {\it {The Two-Loop Six-Gluon MHV Amplitude in Maximally
  Supersymmetric Yang-Mills Theory}},  {\em Phys. Rev. D} {\bf 78} (2008)
  045007, [\href{http://arxiv.org/abs/0803.1465}{{\tt arXiv:0803.1465}}].

\bibitem{CaronHuot:2010ek}
S.~Caron-Huot, {\it {Notes on the scattering amplitude / Wilson loop duality}},
   {\em JHEP} {\bf 07} (2011) 058, [\href{http://arxiv.org/abs/1010.1167}{{\tt
  arXiv:1010.1167}}].

\bibitem{Mason:2010yk}
L.~Mason and D.~Skinner, {\it {The Complete Planar S-matrix of N=4 SYM as a
  Wilson Loop in Twistor Space}},  {\em JHEP} {\bf 12} (2010) 018,
  [\href{http://arxiv.org/abs/1009.2225}{{\tt arXiv:1009.2225}}].

\bibitem{Basso:2013vsa}
B.~Basso, A.~Sever, and P.~Vieira, {\it {Spacetime and Flux Tube S-Matrices at
  Finite Coupling for N=4 Supersymmetric Yang-Mills Theory}},  {\em Phys. Rev.
  Lett.} {\bf 111} (2013), no.~9 091602,
  [\href{http://arxiv.org/abs/1303.1396}{{\tt arXiv:1303.1396}}].

\bibitem{Beisert:2010jr}
N.~Beisert et~al., {\it {Review of AdS/CFT Integrability: An Overview}},  {\em
  Lett. Math. Phys.} {\bf 99} (2012) 3--32,
  [\href{http://arxiv.org/abs/1012.3982}{{\tt arXiv:1012.3982}}].

\bibitem{Alday:2010ku}
L.~F. Alday, D.~Gaiotto, J.~Maldacena, A.~Sever, and P.~Vieira, {\it {An
  Operator Product Expansion for Polygonal null Wilson Loops}},  {\em JHEP}
  {\bf 04} (2011) 088, [\href{http://arxiv.org/abs/1006.2788}{{\tt
  arXiv:1006.2788}}].

\bibitem{CaronHuot:2011kk}
S.~Caron-Huot and S.~He, {\it {Jumpstarting the All-Loop S-Matrix of Planar N=4
  Super Yang-Mills}},  {\em JHEP} {\bf 07} (2012) 174,
  [\href{http://arxiv.org/abs/1112.1060}{{\tt arXiv:1112.1060}}].

\bibitem{Caron-Huot:2011zgw}
S.~Caron-Huot, {\it {Superconformal symmetry and two-loop amplitudes in planar
  N=4 super Yang-Mills}},  {\em JHEP} {\bf 12} (2011) 066,
  [\href{http://arxiv.org/abs/1105.5606}{{\tt arXiv:1105.5606}}].

\bibitem{Li:2021bwg}
Z.~Li and C.~Zhang, {\it {The three-loop MHV octagon from $ \overline{Q} $
  equations}},  {\em JHEP} {\bf 12} (2021) 113,
  [\href{http://arxiv.org/abs/2110.00350}{{\tt arXiv:2110.00350}}].

\bibitem{Caron-Huot:2012awx}
S.~Caron-Huot and K.~J. Larsen, {\it {Uniqueness of two-loop master contours}},
   {\em JHEP} {\bf 10} (2012) 026, [\href{http://arxiv.org/abs/1205.0801}{{\tt
  arXiv:1205.0801}}].

\bibitem{Bourjaily:2017bsb}
J.~L. Bourjaily, A.~J. McLeod, M.~Spradlin, M.~von Hippel, and M.~Wilhelm, {\it
  {Elliptic Double-Box Integrals: Massless Scattering Amplitudes beyond
  Polylogarithms}},  {\em Phys. Rev. Lett.} {\bf 120} (2018), no.~12 121603,
  [\href{http://arxiv.org/abs/1712.02785}{{\tt arXiv:1712.02785}}].

\bibitem{Kristensson:2021ani}
A.~Kristensson, M.~Wilhelm, and C.~Zhang, {\it {Elliptic Double Box and
  Symbology Beyond Polylogarithms}},  {\em Phys. Rev. Lett.} {\bf 127} (2021),
  no.~25 251603, [\href{http://arxiv.org/abs/2106.14902}{{\tt
  arXiv:2106.14902}}].

\bibitem{Bourjaily:2015jna}
J.~L. Bourjaily and J.~Trnka, {\it {Local Integrand Representations of All
  Two-Loop Amplitudes in Planar SYM}},  {\em JHEP} {\bf 08} (2015) 119,
  [\href{http://arxiv.org/abs/1505.05886}{{\tt arXiv:1505.05886}}].

\bibitem{ArkaniHamed:2009dn}
N.~Arkani-Hamed, F.~Cachazo, C.~Cheung, and J.~Kaplan, {\it {A Duality For The
  S Matrix}},  {\em JHEP} {\bf 03} (2010) 020,
  [\href{http://arxiv.org/abs/0907.5418}{{\tt arXiv:0907.5418}}].

\bibitem{Usyukina:1993ch}
N.~I. Usyukina and A.~I. Davydychev, {\it {Exact results for three and four
  point ladder diagrams with an arbitrary number of rungs}},  {\em Phys. Lett.
  B} {\bf 305} (1993) 136--143.

\bibitem{Bourjaily:2013mma}
J.~L. Bourjaily, S.~Caron-Huot, and J.~Trnka, {\it {Dual-Conformal
  Regularization of Infrared Loop Divergences and the Chiral Box Expansion}},
  {\em JHEP} {\bf 01} (2015) 001, [\href{http://arxiv.org/abs/1303.4734}{{\tt
  arXiv:1303.4734}}].

\bibitem{He:2022ctv}
S.~He, Z.~Li, R.~Ma, Z.~Wu, Q.~Yang, and Y.~Zhang, {\it {A study of Feynman
  integrals with uniform transcendental weights and the symbology from dual
  conformal symmetry}},  \href{http://arxiv.org/abs/2206.04609}{{\tt
  arXiv:2206.04609}}.

\bibitem{He:2020uhb}
S.~He and Z.~Li, {\it {A Note on Letters of Yangian Invariants}},  {\em JHEP}
  {\bf 02} (2021) 155, [\href{http://arxiv.org/abs/2007.01574}{{\tt
  arXiv:2007.01574}}].

\bibitem{Mago:2020kmp}
J.~Mago, A.~Schreiber, M.~Spradlin, and A.~Volovich, {\it {Symbol alphabets
  from plabic graphs}},  {\em JHEP} {\bf 10} (2020) 128,
  [\href{http://arxiv.org/abs/2007.00646}{{\tt arXiv:2007.00646}}].

\bibitem{Arkani-Hamed:2019rds}
N.~Arkani-Hamed, T.~Lam, and M.~Spradlin, {\it {Non-perturbative geometries for
  planar $ \mathcal{N} $ = 4 SYM amplitudes}},  {\em JHEP} {\bf 03} (2021) 065,
  [\href{http://arxiv.org/abs/1912.08222}{{\tt arXiv:1912.08222}}].

\bibitem{Drummond:2019cxm}
J.~Drummond, J.~Foster, O.~G\"urdogan, and C.~Kalousios, {\it {Algebraic
  singularities of scattering amplitudes from tropical geometry}},  {\em JHEP}
  {\bf 04} (2021) 002, [\href{http://arxiv.org/abs/1912.08217}{{\tt
  arXiv:1912.08217}}].

\bibitem{Drummond:2019qjk}
J.~Drummond, J.~Foster, O.~G\"urdogan, and C.~Kalousios, {\it {Tropical
  Grassmannians, cluster algebras and scattering amplitudes}},  {\em JHEP} {\bf
  04} (2020) 146, [\href{http://arxiv.org/abs/1907.01053}{{\tt
  arXiv:1907.01053}}].

\bibitem{Drummond:2020kqg}
J.~Drummond, J.~Foster, O.~G\"urdo\u{g}an, and C.~Kalousios, {\it {Tropical
  fans, scattering equations and amplitudes}},  {\em JHEP} {\bf 11} (2021) 071,
  [\href{http://arxiv.org/abs/2002.04624}{{\tt arXiv:2002.04624}}].

\bibitem{Henke:2019hve}
N.~Henke and G.~Papathanasiou, {\it {How tropical are seven- and eight-particle
  amplitudes?}},  {\em JHEP} {\bf 08} (2020) 005,
  [\href{http://arxiv.org/abs/1912.08254}{{\tt arXiv:1912.08254}}].

\bibitem{Henke:2021ity}
N.~Henke and G.~Papathanasiou, {\it {Singularities of eight- and nine-particle
  amplitudes from cluster algebras and tropical geometry}},  {\em JHEP} {\bf
  10} (2021) 007, [\href{http://arxiv.org/abs/2106.01392}{{\tt
  arXiv:2106.01392}}].

\bibitem{He:2021eec}
S.~He, Z.~Li, and Q.~Yang, {\it {Kinematics, cluster algebras and Feynman
  integrals}},  \href{http://arxiv.org/abs/2112.11842}{{\tt arXiv:2112.11842}}.

\bibitem{He:2021esx}
S.~He, Z.~Li, and Q.~Yang, {\it {Notes on cluster algebras and some all-loop
  Feynman integrals}},  {\em JHEP} {\bf 06} (2021) 119,
  [\href{http://arxiv.org/abs/2103.02796}{{\tt arXiv:2103.02796}}].

\bibitem{He:2021non}
S.~He, Z.~Li, and Q.~Yang, {\it {Truncated cluster algebras and Feynman
  integrals with algebraic letters}},  {\em JHEP} {\bf 12} (2021) 110,
  [\href{http://arxiv.org/abs/2106.09314}{{\tt arXiv:2106.09314}}]. [Erratum:
  JHEP 05, 075 (2022)].

\bibitem{Yang:2022gko}
Q.~Yang, {\it {Schubert problems, positivity and symbol letters}},  {\em JHEP}
  {\bf 08} (2022) 168, [\href{http://arxiv.org/abs/2203.16112}{{\tt
  arXiv:2203.16112}}].

\bibitem{He:2020uxy}
S.~He, Z.~Li, Y.~Tang, and Q.~Yang, {\it {The Wilson-loop $d$ log
  representation for Feynman integrals}},  {\em JHEP} {\bf 05} (2021) 052,
  [\href{http://arxiv.org/abs/2012.13094}{{\tt arXiv:2012.13094}}].

\bibitem{He:2020lcu}
S.~He, Z.~Li, Q.~Yang, and C.~Zhang, {\it {Feynman Integrals and Scattering
  Amplitudes from Wilson Loops}},  {\em Phys. Rev. Lett.} {\bf 126} (2021)
  231601, [\href{http://arxiv.org/abs/2012.15042}{{\tt arXiv:2012.15042}}].

\bibitem{Hodges:2009hk}
A.~Hodges, {\it {Eliminating spurious poles from gauge-theoretic amplitudes}},
  {\em JHEP} {\bf 05} (2013) 135, [\href{http://arxiv.org/abs/0905.1473}{{\tt
  arXiv:0905.1473}}].

\bibitem{Drummond:2008vq}
J.~M. Drummond, J.~Henn, G.~P. Korchemsky, and E.~Sokatchev, {\it {Dual
  superconformal symmetry of scattering amplitudes in N=4 super-Yang-Mills
  theory}},  {\em Nucl. Phys.} {\bf B828} (2010) 317--374,
  [\href{http://arxiv.org/abs/0807.1095}{{\tt arXiv:0807.1095}}].

\bibitem{Mason:2009qx}
L.~J. Mason and D.~Skinner, {\it {Dual Superconformal Invariance, Momentum
  Twistors and Grassmannians}},  {\em JHEP} {\bf 11} (2009) 045,
  [\href{http://arxiv.org/abs/0909.0250}{{\tt arXiv:0909.0250}}].

\bibitem{Bern:2005iz}
Z.~Bern, L.~J. Dixon, and V.~A. Smirnov, {\it {Iteration of planar amplitudes
  in maximally supersymmetric Yang-Mills theory at three loops and beyond}},
  {\em Phys. Rev.} {\bf D72} (2005) 085001,
  [\href{http://arxiv.org/abs/hep-th/0505205}{{\tt hep-th/0505205}}].

\bibitem{Alday:2009zm}
L.~F. Alday, J.~M. Henn, J.~Plefka, and T.~Schuster, {\it {Scattering into the
  fifth dimension of N=4 super Yang-Mills}},  {\em JHEP} {\bf 01} (2010) 077,
  [\href{http://arxiv.org/abs/0908.0684}{{\tt arXiv:0908.0684}}].

\bibitem{Henn:2010ir}
J.~M. Henn, S.~G. Naculich, H.~J. Schnitzer, and M.~Spradlin, {\it {More loops
  and legs in Higgs-regulated N=4 SYM amplitudes}},  {\em JHEP} {\bf 08} (2010)
  002, [\href{http://arxiv.org/abs/1004.5381}{{\tt arXiv:1004.5381}}].

\bibitem{Bourjaily:2019vby}
J.~L. Bourjaily, M.~Volk, and M.~Von~Hippel, {\it {Conformally Regulated Direct
  Integration of the Two-Loop Heptagon Remainder}},  {\em JHEP} {\bf 02} (2020)
  095, [\href{http://arxiv.org/abs/1912.05690}{{\tt arXiv:1912.05690}}].

\bibitem{Caron-Huot:2013vda}
S.~Caron-Huot and S.~He, {\it {Three-loop octagons and $n$-gons in maximally
  supersymmetric Yang-Mills theory}},  {\em JHEP} {\bf 08} (2013) 101,
  [\href{http://arxiv.org/abs/1305.2781}{{\tt arXiv:1305.2781}}].

\bibitem{Panzer:2014caa}
E.~Panzer, {\it {Algorithms for the symbolic integration of hyperlogarithms
  with applications to Feynman integrals}},  {\em Comput. Phys. Commun.} {\bf
  188} (2015) 148--166, [\href{http://arxiv.org/abs/1403.3385}{{\tt
  arXiv:1403.3385}}].

\bibitem{Duhr:2019tlz}
C.~Duhr and F.~Dulat, {\it {PolyLogTools — polylogs for the masses}},  {\em
  JHEP} {\bf 08} (2019) 135, [\href{http://arxiv.org/abs/1904.07279}{{\tt
  arXiv:1904.07279}}].

\bibitem{Goncharov:2010jf}
A.~B. Goncharov, M.~Spradlin, C.~Vergu, and A.~Volovich, {\it {Classical
  Polylogarithms for Amplitudes and Wilson Loops}},  {\em Phys. Rev. Lett.}
  {\bf 105} (2010) 151605, [\href{http://arxiv.org/abs/1006.5703}{{\tt
  arXiv:1006.5703}}].

\bibitem{Caron-Huot:2011dec}
S.~Caron-Huot and S.~He, {\it {Jumpstarting the All-Loop S-Matrix of Planar N=4
  Super Yang-Mills}},  {\em JHEP} {\bf 07} (2012) 174,
  [\href{http://arxiv.org/abs/1112.1060}{{\tt arXiv:1112.1060}}].

\bibitem{Gaiotto:2011dt}
D.~Gaiotto, J.~Maldacena, A.~Sever, and P.~Vieira, {\it {Pulling the straps of
  polygons}},  {\em JHEP} {\bf 12} (2011) 011,
  [\href{http://arxiv.org/abs/1102.0062}{{\tt arXiv:1102.0062}}].

\bibitem{He:2021mme}
S.~He, Z.~Li, and Q.~Yang, {\it {Comments on all-loop constraints for
  scattering amplitudes and Feynman integrals}},  {\em JHEP} {\bf 01} (2022)
  073, [\href{http://arxiv.org/abs/2108.07959}{{\tt arXiv:2108.07959}}].

\bibitem{He:2022tph}
S.~He, J.~Liu, Y.~Tang, and Q.~Yang, {\it {The symbology of Feynman integrals
  from twistor geometries}},  \href{http://arxiv.org/abs/2207.13482}{{\tt
  arXiv:2207.13482}}.

\bibitem{Chicherin:2022gky}
D.~Chicherin, J.~Henn, and S.~Zoia, {\it {Anomalous Ward identities for
  on-shell amplitudes at the conformal fixed point}},
  \href{http://arxiv.org/abs/2207.12249}{{\tt arXiv:2207.12249}}.

\bibitem{Corcoran:2021gda}
L.~Corcoran, F.~Loebbert, and J.~Miczajka, {\it {Yangian Ward identities for
  fishnet four-point integrals}},  {\em JHEP} {\bf 04} (2022) 131,
  [\href{http://arxiv.org/abs/2112.06928}{{\tt arXiv:2112.06928}}].

\bibitem{Bourjaily:2019iqr}
J.~L. Bourjaily, E.~Herrmann, C.~Langer, A.~J. McLeod, and J.~Trnka, {\it
  {Prescriptive Unitarity for Non-Planar Six-Particle Amplitudes at Two
  Loops}},  {\em JHEP} {\bf 12} (2019) 073,
  [\href{http://arxiv.org/abs/1909.09131}{{\tt arXiv:1909.09131}}].

\bibitem{Bourjaily:2019gqu}
J.~L. Bourjaily, E.~Herrmann, C.~Langer, A.~J. McLeod, and J.~Trnka, {\it
  {All-Multiplicity Nonplanar Amplitude Integrands in Maximally Supersymmetric
  Yang-Mills Theory at Two Loops}},  {\em Phys. Rev. Lett.} {\bf 124} (2020),
  no.~11 111603, [\href{http://arxiv.org/abs/1911.09106}{{\tt
  arXiv:1911.09106}}].

\bibitem{Caron-Huot:2012sos}
S.~Caron-Huot and Y.-t. Huang, {\it {The two-loop six-point amplitude in ABJM
  theory}},  {\em JHEP} {\bf 03} (2013) 075,
  [\href{http://arxiv.org/abs/1210.4226}{{\tt arXiv:1210.4226}}].

\bibitem{He:2022toappear}
S.~He, Y.-t. Huang, C.-K. Kuo, and Z.~Li, {\it {to appear}}, .

\bibitem{goncharov2005galois}
A.~B. Goncharov et~al., {\it Galois symmetries of fundamental groupoids and
  noncommutative geometry},  {\em Duke Mathematical Journal} {\bf 128} (2005),
  no.~2 209--284.

\bibitem{Goncharov:2005sla}
A.~B. Goncharov, {\it {Galois symmetries of fundamental groupoids and
  noncommutative geometry}},  {\em Duke Math. J.} {\bf 128} (2005) 209,
  [\href{http://arxiv.org/abs/math/0208144}{{\tt math/0208144}}].

\bibitem{Duhr:2011zq}
C.~Duhr, H.~Gangl, and J.~R. Rhodes, {\it {From polygons and symbols to
  polylogarithmic functions}},  {\em JHEP} {\bf 10} (2012) 075,
  [\href{http://arxiv.org/abs/1110.0458}{{\tt arXiv:1110.0458}}].

\end{thebibliography}\endgroup

\end{document}